\documentclass[published]{JHEP3} 

\JHEP{00(2010)000}

\JHEPspecialurl{http://jhep.sissa.it/JOURNAL/JHEP3.tar.gz}

\usepackage{epsfig,multicol,bbm}
\usepackage{epsfig}

\def\sumint{\hbox{$\sum$}\!\!\!\!\!\!\!\int}
\def\square{\vcenter{\vbox{\hrule height.4pt
          \hbox{\vrule width.4pt height8pt
          \kern8pt\vrule width.4pt}\hrule height.4pt}}}

\def\ranglec{\rangle_{\!\!c}}

\newcommand{\beq}{\begin{equation}}
\newcommand{\eeq}{\end{equation}}
\newcommand{\bqa}{\begin{eqnarray}}
\newcommand{\eqa}{\end{eqnarray}}

\newcommand\fverb{\setbox\fverbbox=\hbox\bgroup\verb}
\newcommand\fverbdo{\egroup\medskip\noindent%
			\fbox{\unhbox\fverbbox}\ }
\newcommand\fverbit{\egroup\item[\fbox{\unhbox\fverbbox}]}
\newbox\fverbbox

\title{Three-loop HTL gluon thermodynamics at intermediate coupling}

\author{Jens O. Andersen \\
Department of Physics, Norwegian University of Science
and Technology, H{\o}gskoleringen 5, N-7491 Trondheim, Norway\\
	E-mail: \email{andersen@tf.phys.ntnu.no}}
\author{
Michael Strickland\\
Department of Physics, Gettysburg College, Gettysburg, PA 17325, 
USA and
Frankfurt Institute for Advanced Studies,  
Ruth-Moufang-Str. 1,
D-60438 Frankfurt am Main, Germany\\
	E-mail: \email{mstrickl@gettysburg.edu}}

\author{Nan Su \\
Frankfurt Institute for Advanced Studies,  
Ruth-Moufang-Str. 1,
D-60438 Frankfurt 
am Main, Germany\\
	E-mail: \email{nansu@fias.uni-frankfurt.de}}
\received{\today} 		
\accepted{\today}		

\preprint{\hepth{9912999}}	

\abstract{We calculate the thermodynamic functions of pure-glue QCD 
to three-loop order using
the hard-thermal-loop perturbation theory (HTLpt)
reorganization of finite temperature
quantum field theory.  We show that at three-loop order hard-thermal-loop 
perturbation theory
is compatible with lattice results for the pressure, energy density, and 
entropy down to 
temperatures $T\simeq3\;T_c$.  
Our results suggest that HTLpt provides a systematic framework that can be
used to calculate static and dynamic quantities for temperatures relevant
at LHC.
}

\begin{document} 


The goal of ultrarelativistic heavy-ion collision experiments is to 
generate energy
densities and temperatures high enough to create a 
quark-gluon plasma.  One of the chief theoretical
questions which has emerged in this area is whether it is more appropriate to
describe the state of matter generated during these collisions using 
weakly-coupled
quantum field theory or a strong-coupling approach based on the AdS/CFT 
correspondence.  
Early data from the Relativistic Heavy Ion Collider (RHIC) at 
Brookhaven National Labs 
indicated that the state of matter created there behaved more like a 
fluid than a plasma and that this ``quark-gluon fluid'' is
strongly coupled \cite{rhicexperiment}.  

In the intervening years, however, work on the perturbative side has shown that 
observables like jet quenching  \cite{pert} and elliptic flow 
\cite{Xu:2007jv} can also be described using a perturbative formalism.  Since in
phenomenological applications predictions are complicated by the 
modeling required to describe, for example, initial-state effects, the 
space-time evolution
of the plasma, and hadronization of the plasma, there are significant 
theoretical uncertainties
remaining.  
Therefore, one is hard put to conclude whether the plasma is strongly or 
weakly coupled 
based solely on RHIC data.  To have a cleaner testing ground one can compare 
theoretical 
calculations with results from lattice quantum chromodynamics (QCD).

Looking forward to the upcoming heavy-ion experiments scheduled to take 
place at the 
Large Hadron Collider (LHC) at the European Organization for Nuclear Research
(CERN) it is 
important to know if, at the higher temperatures generated, one expects a 
strongly-coupled (liquid) or 
weakly-coupled (plasma) description to be more appropriate.  At RHIC, 
initial temperatures 
on the order of one to two times the QCD critical temperature,
$T_c \sim 190$ MeV, were obtained.  At LHC, initial temperatures on
the order of $4-5\;T_c$ are expected.  The key question is, will the 
generated matter behave more
like a plasma of quasiparticles at these higher temperatures.

The calculation of thermodynamic functions
using weakly-coupled quantum field theory has a long 
history~
\cite{shur,toimela,AZ-95,parwani,BN-95,BN-96,Andersen,KZ-96,keijo,gynter,all}.  
The QCD 
free energy is known up to order $g^6 \log(g)$; 
however, the resulting weak-coupling
approximations do not converge at phenomenologically relevant couplings. 
For example, simply comparing
the magnitude of low-order contributions to the QCD
free energy with three quark flavors ($N_f=3$), 
one finds that the $g^3$ contribution is smaller than the $g^2$
contribution only for $g \sim 0.9$ ($\alpha_s  \sim 0.07$) which
corresponds to a temperature of $T \sim 10^5 $ GeV $\sim 5 \times 10^5 \, T_c$. 

In Fig~\ref{pertpressure}, we show the weak-coupling expansion for the
pressure of pure-glue QCD
normalized to that of an ideal gas through order $\alpha_s^{5/2}$.
The various approximations oscillate wildy and show no signs of convergence
in the temperature range shown. The bands are obtained by 
varying the renormalization scale $\mu$ by a factor of two around the value
$\mu=2\pi T$ and we use three-loop running of $\alpha_s$
This oscillating behavior is generic
for hot field theories and not specific to QCD.

\FIGURE{\includegraphics[width=10cm]{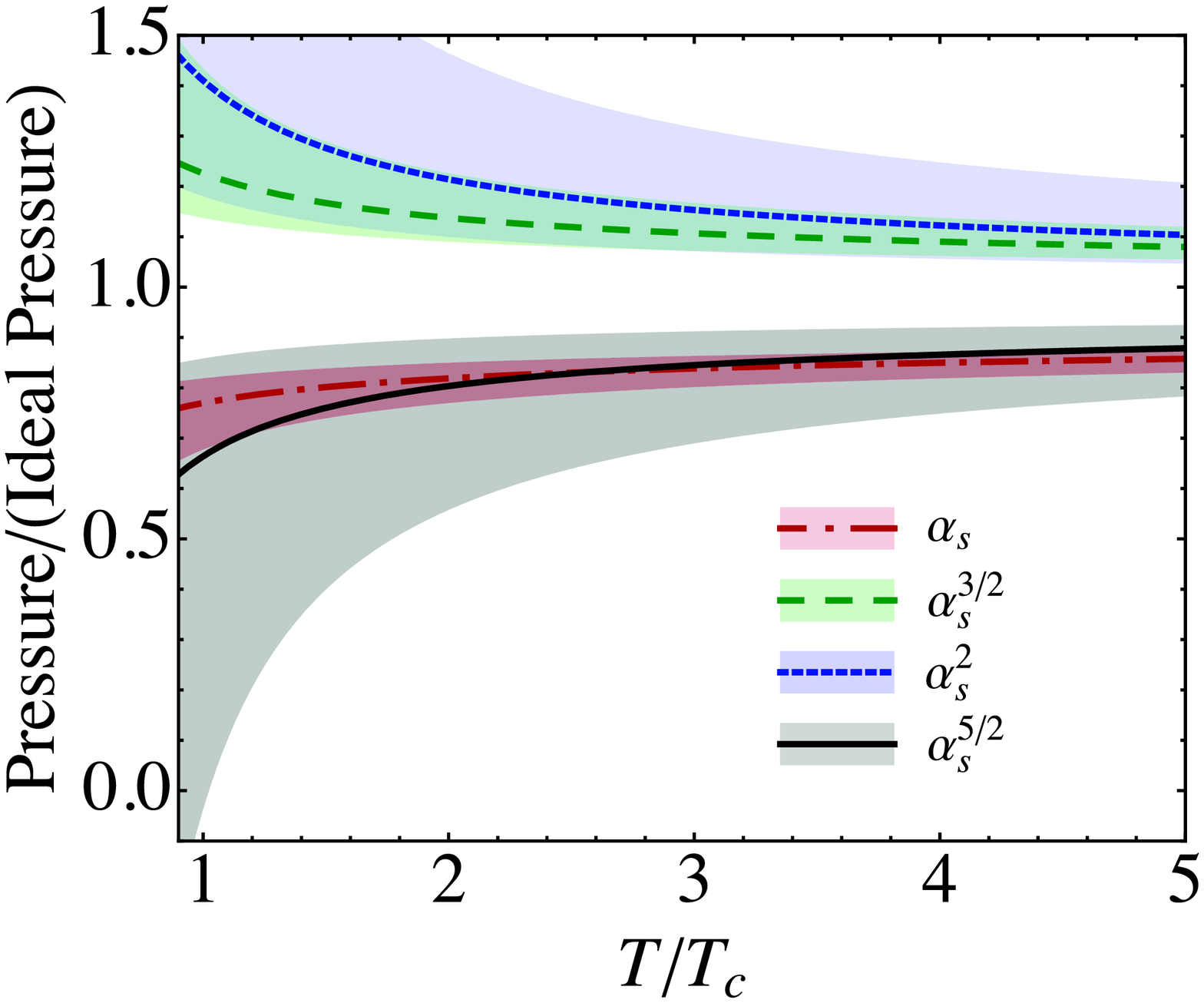}\caption{Weak-coupling expansion for the scaled pressure of pure-glue QCD.
Shaded
bands show the result of varying the renormalization scale $\mu$ by a factor of
two around $\mu = 2 \pi T$.}\label{pertpressure}}

The poor convergence of finite-temperature perturbative expansions of 
thermodynamic functions stems from the fact that at high temperature
the classical solution is not described by massless gluons.  
Instead one must include plasma effects such as the screening
of electric fields and Landau damping 
of excitations
via a self-consistent hard-thermal-loop (HTL)
resummation~\cite{bp}.
There are several ways of systematically 
reorganizing the perturbative expansion \cite{reorg}. Here we will 
present the details of a new NNLO calculation which uses the 
hard-thermal-loop perturbation theory
method \cite{htlpt1,htlpt2,qednnlo,qcdnnlo} 
and compare with previously
obtained LO and NLO results.

The basic idea of the technique is to add and subtract an effective mass term 
from 
the bare Lagrangian and to associate the added piece with the 
free part of the Lagrangian and the subtracted piece with the 
interactions~\cite{prespt,spt}.
However, in gauge theories, one cannot simply add and subtract a local mass
term since this would violate gauge 
invariance~\cite{Braaten:1991gm,Buchmuller:1994qy,Alexanian:1995rp}.
Instead, one adds and subtracts an HTL 
improvement term which modifies the propagators and
vertices self-consistently so that the reorganization is 
manifestly gauge invariant~\cite{Braaten:1991gm}.

In this paper we discuss the calculation of thermodynamic functions of a 
gas of gluons at 
phenomenologically relevant temperatures using hard-thermal-loop perturbation 
theory. We present results at leading 
order (LO),
next-to-leading order (NLO), and next-to-next-to-leading order (NNLO) and 
compare with
available lattice data~\cite{Boyd:1996bx,endrodi} 
for the thermodynamic functions of
SU(3) Yang-Mills theory.  The calculation is based on a reorganization of 
the theory around
hard-thermal-loop (HTL) quasiparticles.  Our results 
indicate that the lattice data at temperatures $T \sim 2 - 3\;T_c$ 
are consistent with the quasiparticle picture.  This is a non-trivial result
since, in this temperature regime, the QCD 
coupling constant is neither infinitesimally weak nor infinitely strong with 
$g \sim 2$, or equivalently $\alpha_s = g^2/(4\pi) \sim 0.3$.  Therefore, 
we 
have a crucial test of the quasiparticle picture in the intermediate coupling 
regime.

\section{HTL perturbation theory}

\label{HTLpt}

The Lagrangian density for pure-glue QCD in Minkowski space is
\bqa
{\cal L}_{\rm QCD}&=&
-{1\over2}{\rm Tr}\left[G_{\mu\nu}G^{\mu\nu}\right]
+{\cal L}_{\rm gf}
+{\cal L}_{\rm gh}
+\Delta{\cal L}_{\rm QCD}\;,
\label{L-QED}
\eqa
%
where the field strength is 
$G^{\mu\nu}=\partial^{\mu}A^{\nu}-\partial^{\nu}A^{\mu}-ig[A^{\mu},A^{\nu}]$.
The ghost term ${\cal L}_{\rm gh}$ depends on the gauge-fixing term
${\cal L}_{\rm gf}$. In this paper we choose the class of covariant gauges
where the gauge-fixing term is
\bqa
{\cal L}_{\rm gf}&=&-{1\over\xi}{\rm Tr}
\left[\left(\partial_{\mu}A^{\mu}\right)^2\right]\;.
\eqa

The perturbative expansion in powers of $g$
generates ultraviolet divergences.
The renormalizability of perturbative QCD guarantees that
all divergences in physical quantities can be removed by
renormalization of the coupling constant $\alpha_s= g^2/4 \pi$.
There is no need for wavefunction renormalization, because
physical quantities are independent of the normalization of
the field.  There is also no need for renormalization of the gauge
parameter, because physical quantities are independent of the
gauge parameter.

Hard-thermal-loop perturbation theory (HTLpt) is a reorganization
of the perturbation
series for thermal QCD. The Lagrangian density is written as
\bqa
{\cal L}= \left({\cal L}_{\rm QCD}
+ {\cal L}_{\rm HTL} \right) \Big|_{g \to \sqrt{\delta} g}
+ \Delta{\cal L}_{\rm HTL}.
\label{L-HTLQCD}
\eqa
The HTL improvement term is
\bqa
{\cal L}_{\rm HTL}=-{1\over2}(1-\delta)m_D^2 {\rm Tr}
\left(G_{\mu\alpha}\left\langle {y^{\alpha}y^{\beta}\over(y\cdot D)^2}
	\right\rangle_{\!\!y}G^{\mu}_{\;\;\beta}\right)
	\, ,
\label{L-HTL}
\eqa
where the covariant derivative is $D^{\mu}=\partial^{\mu}-igA^{\mu}$
and
$y^{\mu}=(1,\hat{{\bf y}})$ is a light-like four-vector,
and $\langle\ldots\rangle_{ y}$
represents the average over the directions
of $\hat{{\bf y}}$.
The term~(\ref{L-HTL}) has the form of the effective Lagrangian
that would be induced by
a rotationally invariant ensemble of charged sources with infinitely high
momentum. The parameter $m_D$ can be identified with the
Debye screening mass.
HTLpt is defined by treating
$\delta$ as a formal expansion parameter.

The HTL perturbation expansion generates ultraviolet divergences.
In perturbative QCD, renormalizability constrains the ultraviolet
divergences to have a form that can be cancelled by the counterterm
Lagrangian $\Delta{\cal L}_{\rm QCD}$.
We will demonstrate that renormalized perturbation theory can be implemented 
by including a counterterm Lagrangian $\Delta{\cal L}_{\rm HTL}$ among 
the interaction terms in (\ref{L-HTLQCD}).
There is no proof that the HTL perturbation expansion is renormalizable,
so the general structure of the ultraviolet divergences is not known;
however, it was shown in previous papers~\cite{htlpt1,htlpt2} 
that it was
possible to renormalize the NLO order HTLpt prediction for the
free energy of QCD using only a vacuum counterterm,
a Debye mass counterterm, and a fermion mass counterterm.  In
this paper we will show that this is also possible at NNLO.
In particular, the only new counterterm we need to introduce is for the
coupling constant $\alpha_s$, which coincides with its perturbative
value giving rise to the standard one-loop running.

We find that the counterterms
necessary to renormalize HTLpt at NNLO are
%
\bqa
\delta\Delta\alpha_s&=&-{11N_c\over12\pi\epsilon}\alpha_s^2\delta^2
+{\cal O}(\delta^3\alpha^3_s)\;,
\label{delalpha}
\\ 
\Delta m_D^2&=&\left(-{11N_c\over12\pi\epsilon}\alpha_s\delta
+{\cal O}(\delta^2\alpha_s^2)
\right)(1-\delta)m_D^2\;,
\label{delmd} \\ 
\Delta{\cal E}_0&=&\left({N_c^2-1\over128\pi^2\epsilon}
+{\cal O}(\delta\alpha_s)
\right)(1-\delta)^2m_D^4\;.
\label{del1e0}
\eqa
We note that the counterterm in Eq.~(\ref{delalpha}) coincides with
the perturbative one-loop running.

Physical observables are calculated in HTLpt
by expanding them in powers of $\delta$,
truncating at some specified order, and then setting $\delta=1$.
This defines a reorganization of the perturbation series
in which the effects of
$m_D^2$ term in~(\ref{L-HTL})
are included to all orders but then systematically subtracted out
at higher orders in perturbation theory
by the $\delta m_D^2$ terms in~(\ref{L-HTL}).
If we set $\delta=1$, the Lagrangian (\ref{L-HTLQCD})
reduces to the QCD Lagrangian (\ref{L-QED}).

If the expansion in $\delta$ could be calculated to all orders
the final result would not depend on $m_D$ when we set $\delta=1$.
However, any truncation of the expansion in $\delta$ produces results
that depend on $m_D$.
Some prescription is required to determine $m_D$ 
as a function of $T$ and $\alpha_s$.
We will discuss several prescriptions in Sec.~VI.
%
%

\section{Diagrams for the thermodynamic potential}
In this section, we list the expressions for the diagrams that contribute
to the thermodynamic potential through order $\delta^2$ in HTL
perturbation theory. The diagrams are shown in Figs.~\ref{fig:dia1},
and \ref{fig:dia2}.  A key to the diagrams is given in Fig.~\ref{keydiagrams}.
The expressions here will be given in Euclidean space; however, in Appendix
\ref{app:rules} we present the HTLpt Feynman rules in Minkowski space.

\FIGURE{\includegraphics[width=9.5cm]{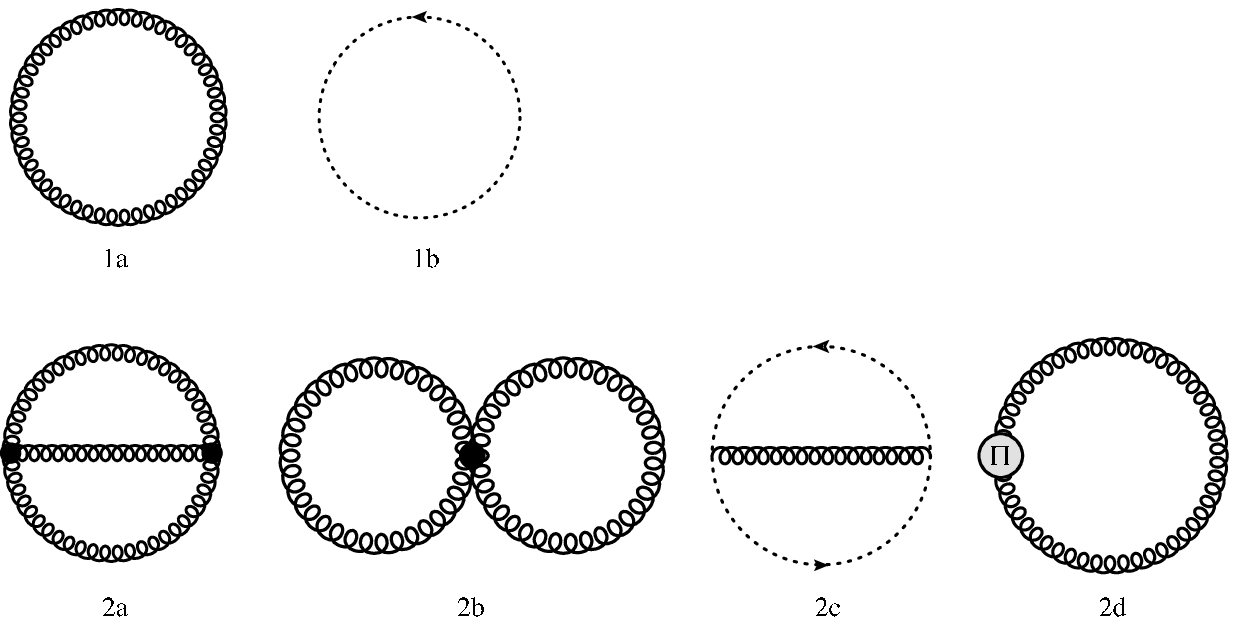}
\caption{Diagrams contributing through NLO in HTLpt.
The spiral lines are gluon propagators and the dotted lines
are ghost propagators.
A circle with a $\Pi$ indicates a gluon self-energy insertion.
All propagators and vertices shown are HTL-resummed propagators and vertices.}
\label{fig:dia1}}

\FIGURE{\includegraphics[width=14.2cm]{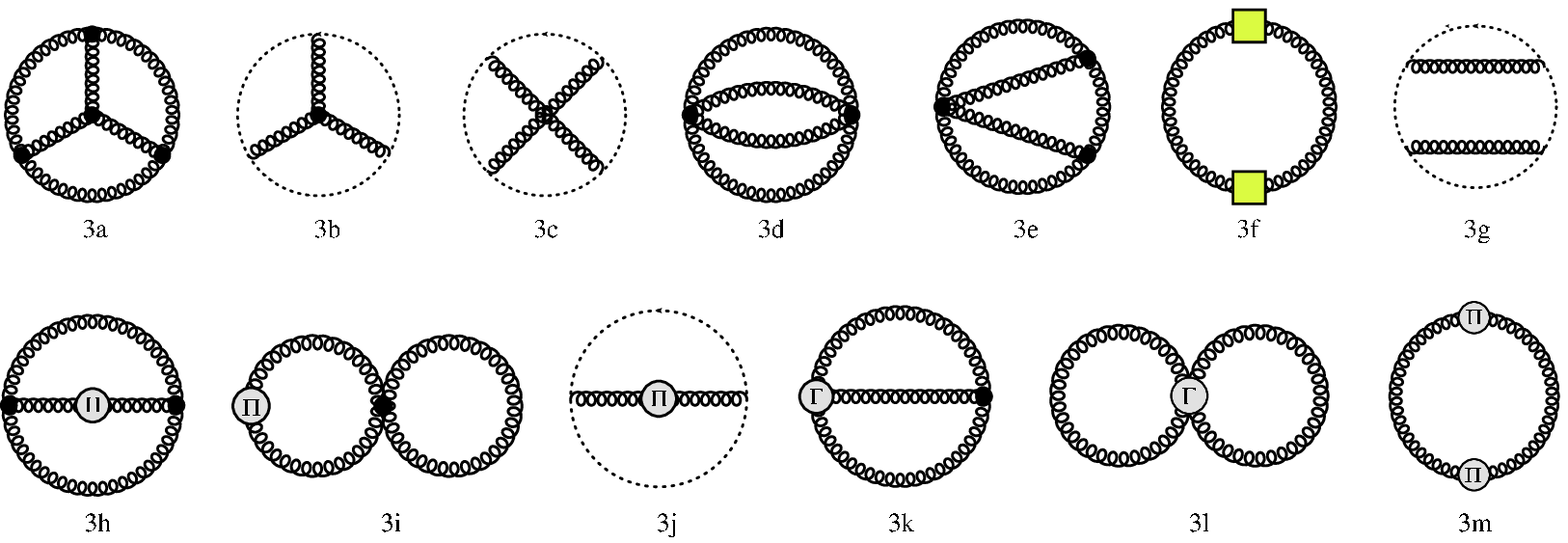}
\caption{Diagrams contributing to NNLO in HTLpt which 
contribute through order $g^5$.
The spiral lines are gluon propagators and the dotted lines
are ghost propagators.
A circle with a $\Pi$ indicates a gluon self-energy insertion 
The propagators are HTL-resummed propagators 
and the black dots 
indicate HTL-resummed vertices. The lettered vertices indicate
that only the HTL correction is included.
The yellow box denotes the insertion of the one-loop self-energy
defined in Fig.~\ref{keydiagrams}.}
\label{fig:dia2}}

\FIGURE{\includegraphics[width=9.2cm]{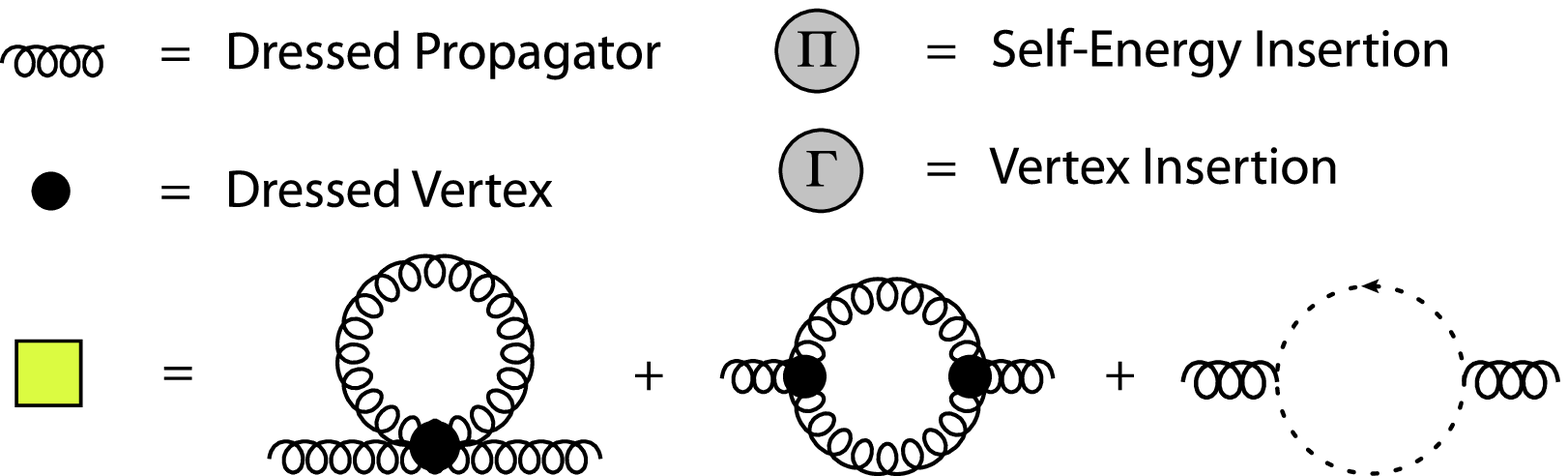}
\caption{Key to the diagrams in Figs.~\ref{fig:dia1} and~\ref{fig:dia2}.}
\label{keydiagrams}}

The thermodynamic potential at leading order in HTL perturbation theory for
QCD is
\bqa
\Omega_{\rm LO}= 
(N_c^2-1)
{\cal F}_{\rm 1a+1b}
+\Delta_0{\cal E}_0\;.
\eqa
Here, ${\cal F}_{\rm 1a+1b}$ is the contribution from the gluon
and ghost diagrams shown on the first line of Fig.~\ref{fig:dia1}
\bqa
\!\!\!\!\!\!\!{\cal F}_{\rm 1a+1b}\!\!&=&\!\!
-{1\over2}\sumint_{P}\!\left\{
(d-1)\log\left[-\Delta_T(P)\right]+\log\Delta_L(P)
\right\}.
\eqa
The transverse and longitudinal HTL propagators
$\Delta_T(P)$ and $\Delta_L(P)$ 
are given in (\ref{Delta-T}) and (\ref{Delta-L}).
The leading-order vacuum counterterm $\Delta_0{\cal E}_0$ was determined
in Ref.~\cite{htlpt1}:
\bqa
\Delta_0{\cal E}_0&=&{N_c^2-1\over128\pi^2\epsilon} m_D^4 \;.
\label{lovac}
\eqa

The thermodynamic potential at 
NLO in HTL perturbation theory
can be written as
\bqa\nonumber
\Omega_{\rm NLO}&=&\Omega_{\rm LO}
 + (N_c^2-1)[{\cal F}_{\rm 2a}+{\cal F}_{\rm 2b}+{\cal F}_{\rm 2c}
+{\cal F}_{\rm 2d}
]
\\&&
+\Delta_1{\cal E}_0
+\Delta_1 m_D^2{\partial\over\partial m_D^2}
\Omega_{\rm LO}
\;,
\label{OmegaNLO}
\eqa
where $\Delta_1{\cal E}_0$ and $\Delta_1m_D^2$
are the terms of order
$\delta$ in the vacuum energy density and mass counterterms:
\bqa
\label{del111}
\Delta_1{\cal E}_0&=&-{N_c^2-1\over64\pi^2\epsilon}m_D^4\;, \\
\Delta_1m_D^2&=&-{11N_c\over12\pi\epsilon}\alpha_s m_D^2\;.
\label{del333}
\eqa
\\
The contributions from the two-loop diagrams
with the three-gluon and four-gluon
vertices are
\bqa\nonumber
{\cal F}_{\rm 2a}
&=&{N_c\over12}g^2\sumint_{PQ}
\Gamma^{\mu\lambda\rho}(P,Q,R)\Gamma^{\nu\sigma\tau}(P,Q,R)
\Delta^{\mu\nu}(P) \\
&&
\times
\Delta^{\lambda\sigma}(Q)\Delta^{\rho\tau}(R)\;,\\
{\cal F}_{\rm 2b} \nonumber
&=&{N_c\over8}g^2\sumint_{PQ}
\Gamma^{\mu\nu,\lambda\sigma}(P,-P,Q,-Q)
\Delta^{\mu\nu}(P)
\\
&&\times\Delta^{\lambda\sigma}(Q)\;,
\eqa
where $R=Q-P$.
For the corresponding diagrams, 
see the second line of Fig.~\ref{fig:dia1}.

The contribution from the ghost diagram is
\bqa
{\cal F}_{\rm 2c}&=&
{N_c\over2}g^2
\sumint_{PQ}
{1\over Q^2}{1\over R^2}Q^{\mu}R^{\nu}\Delta^{\mu\nu}(P)\;.
\eqa

The contribution from the HTL gluon counterterm diagram 
with a single gluon self-energy insertion is
\bqa 
{\cal F}_{\rm 2d}&=&
{1\over2}\sumint_{P}\Pi^{\mu\nu}(P)\Delta^{\mu\nu}(P)\;.
\eqa


The thermodynamic potential at 
NNLO in HTL perturbation theory
can be written as

\bqa\nonumber
\Omega_{\rm NNLO}&=&\Omega_{\rm NLO}
+(N_c^2-1)\left[
{\cal F}_{\rm 3a}+{\cal F}_{\rm 3b}+{\cal F}_{\rm 3c}
+{\cal F}_{\rm 3d}+{\cal F}_{\rm 3e}
+{\cal F}_{\rm 3f}+{\cal F}_{\rm 3g}+{\cal F}_{\rm 3h}+{\cal F}_{\rm 3i}
+{\cal F}_{\rm 3j}+{\cal F}_{\rm 3k}
\right.\\ &&\nonumber\left.
+{\cal F}_{\rm 3l}
+{\cal F}_{\rm 3m}
\right]
+\Delta_2{\cal E}_0
+\Delta_2 m_D^2{\partial\over\partial m_D^2}
\Omega_{\rm LO}
+\Delta_1 m_D^2{\partial\over\partial m_D^2}
\Omega_{\rm NLO}
\\ &&
+{1\over2}\left({\partial^2\over(\partial m_D^2)^2}
\Omega_{\rm LO}\right)\left(\Delta_1m_D^2\right)^2
+(N_c^2-1){{\cal F}_{\rm 2a+2b+2c}\over\alpha_s}\Delta_1\alpha_s
\;.
\label{OmegaNNLO}
\eqa
where $\Delta_1\alpha_s$, $\Delta_2m_D^2$, and $\Delta_2{\cal E}_0$
are the terms of order
$\delta^2$ in the coupling constant, mass, and vacuum energy density
counterterms:
\bqa
\Delta_1\alpha_s&=&-{11N_c\over12\pi\epsilon}\alpha_s^2\;,
\label{delalpha2}
\\ 
\Delta_2m_D^2&=&{11N_c\over12\pi\epsilon}\alpha_sm_D^2\;,
\label{delmd2} \\ 
\Delta_2{\cal E}_0&=&{N_c^2-1\over128\pi^2\epsilon}m_D^4\;.
\label{del1e2}
\eqa

The contributions from the three-loop diagrams are given by

\bqa\nonumber
{\cal F}_{\rm 3a}&=&
{N_c^2\over24}g^4\sumint_{PQR}
\Gamma^{\alpha\beta\gamma}(P,Q,-P-Q)\Delta^{\alpha\theta}(P)
\Delta^{\beta\mu}(Q)\Delta^{\gamma\sigma}(P+Q)
\Gamma^{\mu\nu\delta}(-Q,-R,Q+R)
\Delta^{\pi\nu}(R)
\\ &&\times
\Delta^{\delta\lambda}(Q+R)
\Gamma^{\sigma{\lambda}\rho}(P+Q,-Q-R,R-P)
\Delta^{\rho{\phi}}(R-P)\Gamma^{\theta\phi\pi}(-P,P-R,R)\;, 
\\
\nonumber
{\cal F}_{\rm 3b}&=&
{N_c^2\over3}g^4
\sumint_{PQR}{{R}^{\alpha}(P+Q+R)^{\beta}(P+R)^{\gamma}\over R^2(P+R)^2(P+Q+R)^2}
\Gamma^{\mu\lambda\nu}(-P,-Q,P+Q)
\Delta^{\alpha\mu}(P)
\\ &&\times
\Delta^{\beta\nu}(P+Q)\Delta^{\gamma\lambda}(Q)\;,
\\
{\cal F}_{\rm 3c}&=&
-{N_c^2\over4}g^4
\sumint_{PQR}{(Q+R)^{\alpha}(R-P)^{\beta}(Q+R-P)^{\mu}R^{\nu}\over R^2(Q+R)^2(Q+R-P)^2(R-P)^2}
\Delta^{\alpha\beta}(P)\Delta^{\mu\nu}(Q)\;,
\\
{\cal F}_{\rm 3d}&=&
{N_c^2\over48}
\sumint_{PQR}
\Gamma^{\alpha\beta,\mu\nu}(P,Q,R,S)
\Gamma^{\gamma\delta,\sigma\lambda}(P,Q,R,S)
\Delta^{\alpha\gamma}(P)\Delta^{\beta\delta}(Q)
\Delta^{\mu\sigma}(R)\Delta^{\nu\lambda}(S)\;,
\\ \nonumber
{\cal F}_{\rm 3e}&=&
{-}{N_c^2\over{4}}
\sumint_{PQR}\Gamma^{\alpha\mu,\gamma\sigma}(P,Q,R,S)\Delta^{\alpha\beta}(P)
\Delta^{\mu\nu}(Q)\Delta^{\gamma\delta}(R)
\Delta^{\sigma\phi}(S)\Delta^{\theta\lambda}(P+Q)
\\ &&\times
\Gamma^{\beta\nu\theta}(-P,-Q,P+Q)
\Gamma^{\lambda\delta\phi}({ -P-Q,-R,-S)}\;,
\\
{\cal F}_{\rm 3f}&=&
\sumint_{P}\bar{\Pi}^{\mu\nu}(P)\Delta^{\nu\alpha}(P)
\bar{\Pi}^{\alpha\beta}(P)\Delta^{\beta\mu}(P)\;,
\\
{\cal F}_{\rm 3g}&=&
{-}{N_c^2\over2}g^4
\sumint_{PQR}
{P^{\alpha}(P+Q)^{\mu}P^{\nu}(P+R)^{\beta}\over P^4(P+Q)^2(P+R)^2}
\Delta^{\mu\nu}(Q)\Delta^{\alpha\beta}(R)\;,
\eqa
where $S=-(P+Q+R)$ and $\bar{\Pi}^{\mu\nu}(P)$ is the one-loop 
gluon self-energy 
defined by the yellow box in Fig.~\ref{keydiagrams}.
\bqa\nonumber
\bar{\Pi}^{\mu\nu}(P)
&=&{1\over2}N_cg^2\sumint_Q
\Gamma^{\mu\nu,\alpha\beta}(P,-P,Q,-Q)\Delta^{\alpha\beta}(Q)
+{1\over2}N_cg^2\sumint_Q\Gamma^{\mu\alpha\beta}(P,Q, -P-Q)
\Delta^{\alpha\gamma}(Q)
\\&&
\times\Gamma^{\nu\gamma\delta}(P,Q,-P-Q)\Delta^{\beta}
\delta(R)
+N_cg^2\sumint_Q{Q^{\mu}(P+Q)^{\nu}\over Q^2(P+Q)^2}\;.
\eqa
The contributions from the two-loop diagrams with a single self-energy
insertion are
\bqa
{\cal F}_{\rm 3h}&=&
-{N_c\over4}g^2
\sumint_{PQ}\Gamma^{\alpha\mu\nu}(P,Q,R)\Gamma^{\beta\gamma\delta}(P,Q,R)
\Delta^{\alpha\sigma}(P)\Pi^{\sigma\lambda}(P)\Delta^{\lambda\beta}(P)
\Delta^{\mu\gamma}(Q)\Delta^{\nu\delta}(R)\;,\\
{\cal F}_{\rm 3i}&=&
-{N_c\over4}g^2\sumint_{PQ}\Gamma^{\alpha\beta,\mu\nu}(P,-P,Q,-Q)
\Delta^{\alpha\gamma}(P)
\Pi^{\gamma\delta}(P)
\Delta^{\delta\beta}(P)\Delta^{\mu\nu}(Q)\;,\\
{\cal F}_{\rm 3j}&=&
-{N_c\over2}g^2
\sumint_{PQ}
{P^{\alpha}(P+Q)^{\beta}\over P^2(P+Q)^2}
\Delta^{\alpha\mu}(Q)\Pi^{\mu\nu}(Q)\Delta^{\nu\beta}(Q)\;,
\eqa
where $R=Q-P$.

The two-loop diagrams with a subtracted vertex is
\bqa\nonumber
{\cal F}_{\rm 3k}
&=&{N_c\over6}g^2m_D^2\sumint_{PQ}
{\cal T}^{\mu\lambda\rho}(P,Q,R)\Gamma^{\nu\sigma\tau}(P,Q,R)
\Delta^{\mu\nu}(P) \\&&\times
\Delta^{\lambda\sigma}(Q)\Delta^{\rho\tau}(R)\;,\\
{\cal F}_{\rm 3l} \nonumber
&=&{N_c\over8}g^2m_D^2\sumint_{PQ}
{\cal T}^{\mu\nu,\lambda\sigma}(P,-P,Q,-Q)
\Delta^{\mu\nu}(P)
\\&&\times\Delta^{\lambda\sigma}(Q)\;,
\eqa
where $R=Q-P$.

The contribution from the HTL gluon counterterm diagram 
with two gluon self-energy insertions is
\bqa
{\cal F}_{\rm 3m}&=&
-{1\over4}\sumint_{P}\Pi^{\mu\nu}(P)\Delta^{\nu\alpha}(P)\Pi^{\alpha\beta}(P)
\Delta^{\beta\mu}(P)\;.
\eqa

\section{Expansion in $m_D/T$}
In the papers~\cite{htlpt1,htlpt2,qednnlo,qednnlo1,qednnlo2}, 
the free energy was 
reduced to scalar sum-integrals. It was clear that evaluating these
scalar sum-integrals exactly was intractable and the
sum-integrals were calculated approximately 
by expanding them in powers of $m_D/T$.  
We will follow the same strategy in this paper and 
carry out the 
expansion to high enough order  to include all terms 
through  order $g^5$ if $m_D$ is taken to be of order $g$.

The free energy can be divided into contributions from hard and soft momenta.
In the one-loop diagrams, the contributions are either hard $(h)$ or soft 
$(s)$, 
while at the two-loop level, there are hard-hard $(hh)$, hard-soft $(hs)$, 
and soft-soft $(ss)$ contributions.
At three loops there are hard-hard-hard $(hhh)$, hard-hard-soft $(hhs)$, 
hard-soft-soft $(hss)$, and soft-soft-soft $(sss)$
contributions. 

\subsection{Leading order}

\subsubsection{Hard contribution}
For hard momenta, the self-energies are suppressed by $m_D/T$
relative to the propagators, so we can expand in powers
of $\Pi_T(P)$ and $\Pi_L(P)$.

For the one-loop graphs $\rm (1a)$ and $\rm (1b)$, 
we need to expand to second order in $m^2_D$:

\bqa\nonumber
{\cal F}_{\rm 1a+1b}^{(h)}&=&
{1\over2}(d-1)\sumint_P\log\left(P^2\right)+{1\over2}
m_D^2\sumint_P{1\over P^2}
\\ && \nonumber
-{1\over4(d-1)}m_D^4\sumint_P\left[
{1\over P^4}-2{1\over p^2P^2}-2d{1\over p^4}{\cal T}_P
+2{1\over p^2P^2}{\cal T}_P
+d{1\over p^4}\left({\cal T}_P\right)^2
\right]
\\ \nonumber
&=&
- {\pi^2 \over 45} T^4
+ {1 \over 24} \left[ 1
        + \left( 2 + 2{\zeta'(-1) \over \zeta(-1)} \right) \epsilon \right]
\left( {\mu \over 4 \pi T} \right)^{2\epsilon} m_D^2 T^2
\\&& 
- {1 \over 128 \pi^2}
\left( {1 \over \epsilon} - 7 + 2 \gamma_E + {2 \pi^2\over 3} 
\right)
\left( {\mu \over 4 \pi T} \right)^{2\epsilon} m_D^4 \;.
\label{Flo-h}
\eqa

\subsubsection{Soft contribution}
The soft contribution
in the diagrams $(1a+1b)$
arises from the $P_0=0$ term in the sum-integral.
At soft momentum $P=(0,{\bf p})$, the HTL self-energy functions
reduce to $\Pi_T(P) = 0$ and $\Pi_L(P) = m_D^2$.
The transverse term vanishes in dimensional regularization
because there is no momentum scale in the integral over ${\bf p}$.
Thus the soft contributions come from the longitudinal term only and read
\bqa\nonumber
{\cal F}^{(s)}_{\rm 1a+1b}
&=&{1\over2}T\int_p\log\left(p^2+m_D^2\right)\\
&=&- {m_D^3T\over12\pi}
\left( {\mu \over 2 m} \right)^{2 \epsilon}\left[
1+{8\over3}\epsilon
\right]\;.
\label{count11}
\eqa

We have kept the order-$\epsilon$ in the $m_D^2$ and $m_D^3$  
terms, respectively in Eqs.~(\ref{Flo-h})
and~(\ref{count11}) since they contribute in the counterterms at 
next-to-leading order.

\subsection{Next-to-leading order}

\subsubsection{Hard contribution}
The one-loop graph with a gluon self-energy insertion $\rm (2d)$ 
has an explicit factor of $m_D^2$ and so we need only
to expand the sum-integal to first order in $m_D^2$:

\bqa\nonumber
{\cal F}_{\rm 2d}^{(h)}\!\!&=&\!\!-{1\over2}
m_D^2\sumint_P{1\over P^2}
+{1\over2(d-1)}m_D^4\sumint_P\left[
{1\over P^4}-2{1\over p^2P^2}-2d{1\over p^4}{\cal T}_P
+2{1\over p^2P^2}{\cal T}_P
+d{1\over p^4}\left({\cal T}_P\right)^2
\right]\\
\!\!&=&\!\!
-{1 \over 24} \left[ 1
        + \left( 2 + 2{\zeta'(-1) \over \zeta(-1)} \right) \epsilon \right]
\left( {\mu \over 4 \pi T} \right)^{2\epsilon} m_D^2 T^2
+ {1 \over 64 \pi^2}
\left( {1 \over \epsilon} - 7 + 2 \gamma_E + {2 \pi^2\over 3} 
\right)
\left( {\mu \over 4 \pi T} \right)^{2\epsilon} m_D^4 \;.
\label{ct1}
\eqa
\subsubsection{Soft contribution}
The soft contribution from $\rm (2d)$ arises from the $P_0=0$ term
in the sum-integral. Only the longitudinal part $\Pi_L(P)$ of the self-energy
contributes and reads
\bqa\nonumber
{\cal F}^{(s)}_{\rm 2d}&=&
-{1\over2}m_D^2T\int_p{1\over p^2+m_D^2}
\\
&=&
{m^3_DT\over 8\pi} \left( {\mu \over 2 m_D} \right)^{2 \epsilon}
\left[1 + 2 \epsilon 
\right]\;.
\label{count12}
\eqa

\subsubsection{$(hh)$ contributions}

For hard momenta, the self-energies are suppressed by $m_D/T$
relative to the propagators, so we can expand in powers
of $\Pi_T$ and $\Pi_L$.
The two-loop contribution was calculated in Ref.~\cite{htlpt2} 
and reads
\bqa
\nonumber
{\cal F}^{(hh)}_{\rm 2a+2b+2c}&=&
{N_c\over4}g^2(d-1)^2\sumint_{PQ}\left[
{1\over P^2}{1\over Q^2}\right]+
{N_c\over4}g^2m_D^2\sumint_{PQ}\left[
-2(d-1){1\over P^2}{1\over Q^4}+2(d-2){1\over P^2}{1\over q^2Q^2}
\right.\\ \nonumber
&&\left.
+(d+2){1\over Q^2R^2r^2}-2d{P\cdot Q\over P^2Q^2r^4}-4d{q^2\over P^2Q^2r^4}
+4{q^2\over P^2Q^2r^2R^2}-2(d-1){1\over P^2}{1\over q^2Q^2}{\cal T}_Q
\right. \\&&
-(d+1){1\over P^2Q^2r^2}{\cal T}_R
\left.
+4d{q^2\over P^2Q^2r^4}{\cal T}_R
+2d{P\cdot Q\over P^2Q^2r^4}{\cal T}_R
\right]\;.
\label{hh2loop}
\eqa
Using the expressions for the sum-integrals listed in Appendix B, we obtain
\bqa\nonumber
{\cal F}^{(hh)}_{\rm 2a+2b+2c}&=&
{\pi^2\over12}{N_c\alpha_s\over3\pi}
\left[
1+\left(2+4{\zeta^{\prime}(-1)\over\zeta(-1)}\right)\epsilon
\right]\left({\mu\over4\pi T}\right)^{4\epsilon}T^4
\\ &&
-{7\over96}\left[
{1\over\epsilon}+4.621
\right]{N_c\alpha_s\over3\pi}\left({\mu\over4\pi T}\right)^{4\epsilon}
m^2_DT^2\;.
\label{hhcount}
\eqa
\subsubsection{$(hs)$ contributions}
In the $(hs)$ region, the momentum $P$ is soft. 
The momenta $Q$ and $R$ are always hard. The function that multiplies 
the soft propagator $\Delta_T(0,{\bf p})$, $\Delta_L(0,{\bf p})$
or $\Delta_X(0,{\bf p})$
can be expanded in powers of the soft momentum ${\bf p}$. In the case
of $\Delta_T(0,{\bf p})$, the resulting integrals over ${\bf p}$
have no scale and they vanish in dimensional regularization.
The integration measure $\int_{p}$ scales like $m_D^3$,
the soft propagators $\Delta_L(0,{\bf p})$
and $\Delta_X(0,{\bf p})$ scale like $1/m_D^2$,
and every power of $p$ in the numerator scales like $m_D$.
The two-loop contribution was calculated in Ref.~\cite{htlpt2} 
and reads
\bqa\nonumber
{\cal F}_{\rm 2a+2b+2c}^{(hs)}&=&{N_c\over2}g^2T\int_{p}{1\over p^2+m_D^2}
\sumint_Q\left[-(d-1){1\over Q^2}+2(d-1){q^2\over Q^4}\right]
+N_cg^2m_D^2T\int_{p}{1\over p^2+m_D^2}
\\ &&
\sumint_Q\left[-(d-4){1\over Q^4}
+{(d-1)(d+2)\over d}{q^2\over Q^6}
-{4(d-1)\over d}{q^4\over Q^8}
\right]\;.
\eqa

In order to facilitate the calculations, it proves useful to isolate
the terms that are specific to HTL perturbation theory. After integrating by 
parts and using the results from Zhai and Kastening~\cite{KZ-96}, 
we can write
\bqa\nonumber
{\cal F}_{\rm 2a+2b+2c}^{(hs)}&=&{N_c\over2}g^2T(d-1)^2
\int_{p}{1\over p^2+m_D^2}
\sumint_Q{1\over Q^2}+
{N_c\over12}[d^2-5d+16]g^2Tm_D^2
\int_{p}{1\over p^2+m_D^2}
\sumint_Q{1\over Q^4}
\\ &&
-
{N_c\over2}(d-5)g^2Tm_D^2
\int_{p}{1\over p^2+m_D^2}
\sumint_Q{1\over Q^4}\;.
\eqa
Using the expressions for the integrals and sum-integrals in Appendices
B and C,
we obtain
\bqa\nonumber
{\cal F}^{(hs)}_{\rm 2a+2b+2c}
&=&
-{\pi\over2}{N_c\alpha_s\over3\pi}m_DT^3\left[
1+\left(2+2{\zeta^{\prime}(-1)\over\zeta(-1)}\right)\epsilon
\right]\left({\mu\over4\pi T}\right)^{2\epsilon}
\left({\mu\over2m_D}\right)^{2\epsilon}
\\ &&
-{11\over32\pi}\left({1\over\epsilon}+{27\over11}+2\gamma_E\right)
{N_c\alpha_s\over3\pi}
\left({\mu\over4\pi T}\right)^{2\epsilon}
\left({\mu\over2m_D}\right)^{2\epsilon}m_D^3T\;.
\label{hscount}
\eqa

\subsubsection{$(ss)$ contribution}
The $(ss)$ contribution to the free energy is given by 
a two-loop calculation in electrostatic QCD (EQCD) in 
three dimensions. This calculation was
carried out in Ref.~\cite{BN-96} by Braaten and Nieto. Alternatively,
one can isolate the $(ss)$ contributions from the two-loop diagrams
which were calculated by Arnold and Zhai in Ref.~\cite{AZ-95}.
In Ref.~\cite{htlpt2}, this contribution was calculated and agrees with earlier
results. One finds
\bqa\nonumber
{\cal F}^{(ss)}_{\rm 2a+2b+2c}&=&
{1\over4}N_cg^2T^2\int_{pq}{p^2+4m_D^2\over p^2(q^2+m_D^2)(r^2+m_D^2)}
\\
&=&
{3\over16}
\left[{1\over\epsilon}+3\right]{N_c\alpha_s\over3\pi}
\left({\mu\over2m_D}\right)^{4\epsilon}m_D^2T^2
\;.
\eqa
We have kept the order $\epsilon$ in 
terms Eqs.~(\ref{ct1}),~(\ref{count12}),~(\ref{hhcount}), 
and~(\ref{hscount}) 
since they contribute in the counterterms at 
NNLO.

\subsection{Next-to-next-to-leading order}
\subsubsection{Hard contribution}
The one-loop graph with two gluon self-energy insertions $\rm (3m)$
is proportional to $m_D^4$ and so
must be expanded to zeroth order in $m_D^2$
\bqa\nonumber
{\cal F}_{\rm 3m}^{(h)}&=&
-{1\over4(d-1)}m_D^4\sumint_P\left[
{1\over P^4}-2{1\over p^2P^2}-2d{1\over p^4}{\cal T}_P
+2{1\over p^2P^2}{\cal T}_P
+d{1\over p^4}\left({\cal T}_P\right)^2
\right]\\
&=&
-{1 \over 128\pi^2}
\left( {1 \over \epsilon} - 7 + 2 \gamma_E + {2 \pi^2\over 3} 
\right)
\left( {\mu \over 4 \pi T} \right)^{2\epsilon} m_D^4 \,.
\label{ct2}
\eqa
\subsubsection{Soft contribution}
The soft contribution from $\rm (3m)$ arises from the $P_0=0$ term
in the sum-integral. Only the longitudinal part $\Pi_L(P)$ of the self-energy
contributes and reads
\bqa\nonumber
{\cal F}^{(s)}_{\rm 3m}&=& - {1\over4}m_D^4T\int_p{1\over(p^2+m_D^2)^2}
\\
&=& - {m^3_DT\over32\pi}
\;.
\eqa

\subsubsection{$(hh)$ contributions}
We also need the $(hh)$ contribution from the diagrams $\rm (3h)-(3l)$. 
We calculate their contributions by expanding the two-loop diagrams
$\rm (2a)-(2c)$ to first order in $m_D^2$. This yields
\bqa\nonumber
{\cal F}^{(hh)}_{\rm 3h-3l}
&=&-{N_c\over4}g^2m_D^2\sumint_{PQ}\left[
-2(d-1){1\over P^2}{1\over Q^4}+2(d-2){1\over P^2}{1\over q^2Q^2}
+(d+2){1\over Q^2R^2r^2}
\right.\\ \nonumber&&\left.
-2d{P\cdot Q\over P^2Q^2r^4}-4d{q^2\over P^2Q^2r^4}
+4{q^2\over P^2Q^2r^2R^2}
-2(d-1){1\over P^2}{1\over q^2Q^2}{\cal T}_Q
\right. \\ \nonumber&&\left.
-(d+1){1\over P^2Q^2r^2}{\cal T}_R
+4d{q^2\over P^2Q^2r^4}{\cal T}_R
+2d{P\cdot Q\over P^2Q^2r^4}{\cal T}_R
\right]\\
&=&
{7\over96}\left[
{1\over\epsilon}+4.621
\right]{N_c\alpha_s\over3\pi}\left({\mu\over4\pi T}\right)^{4\epsilon}
m^2_DT^2\;.
\label{hh2loopself}
\eqa
\subsubsection{$(hs)$ contributions}
We also need the $(hs)$ contribution from the diagrams $\rm (3h)-(3l)$. 
Again we calculate their contributions by expanding the two-loop diagrams
$\rm (2a)-(2c)$ to first order in $m_D^2$. This yields
\bqa\nonumber
{\cal F}_{\rm 3h-3l}^{(hs)}
&=&{N_c\over2}g^2(d-1)^2m_D^2T\int_{p}{1\over(p^2+m_D^2)^2}
\sumint_Q{1\over Q^2} \nonumber \\
&& \hspace{5mm} 
-{N_c\over12}g^2m_D^2T\left[d^2-5d+16\right]\int_{p}{p^2\over(p^2+m_D^2)^2}
\sumint_Q{1\over Q^4} \nonumber
\\ && \hspace{10mm} 
+{N_c\over2}g^2(d-5)m_D^2T\int_{p}{p^2\over(p^2+m_D^2)^2}
\sumint_Q{1\over Q^4}\;.
\eqa
This yields
\bqa\nonumber
{\cal F}_{\rm 3h-3l}^{(hs)}&=&
{\pi\over4}{N_c\alpha_s\over3\pi}m_DT^3
+{33\over64\pi}\left({1\over\epsilon}+{59\over33}+2\gamma_E\right)
{N_c\alpha_s\over3\pi}
\left({\mu\over4\pi T}\right)^{2\epsilon}
\left({\mu\over2m_D}\right)^{2\epsilon}m_D^3T\;.
\\ &&
\eqa

\subsubsection{$(ss)$ contribution}

The $(ss)$ contribution from the two-loop diagrams with a single
self-energy insertion or vertex correction 
can be obtained by expanding
the two-loop result in powers of $m_D^2$. This yields
\bqa\nonumber
{\cal F}^{(ss)}_{\rm 3h-3l}&=&-{1\over4}N_cg^2m_D^2T^2
\int_{pq}
\left[
{4\over p^2(q^2+m^2_D)(r^2+m_D^2)}
-{2(p^2+4m_D^2)\over p^2(q^2+m^2_D)^2(r^2+m_D^2)}
\right]
\\
&=&
-{3\over16}
\left[{1\over\epsilon}+1\right]{N_c\alpha_s\over3\pi}
\left({\mu\over2m_D}\right)^{4\epsilon}m_D^2T^2\;.
\eqa
We have verified this by explicitly calculating the relevant diagrams.

\subsubsection{$(hhh)$ contribution}
If all the three loop momenta are hard, we can obtain the 
$m_D/T$ expansion simply by expanding in powers of $m_D^2$. To obtain the 
expansion through order $g^5$, we can use bare propagators and 
vertices. The contributions from the three-loop diagrams were first calculated
by Arnold and Zhai in Ref.~\cite{AZ-95}, and later by 
Braaten and Nieto~\cite{BN-96}.
One finds

\bqa\nonumber
{\cal F}^{(hhh)}_{\rm 3a-3g}&=&
{N_c^2\over4}g^4(d-1)^2
\sumint_{PQR}\Biggl[
-(d-5){1\over P^2Q^2R^4}
-{1\over2}{1\over P^2Q^2R^2(P+Q+R)^2} \\
&& \hspace{47mm} -{(P-Q)^4\over P^2Q^2R^4(Q-R)^2(R-P)^2}
\Biggr] \nonumber \\
&=&-{25\pi^2\over48}\left({N_c\alpha_s\over3\pi}\right)^2
\left[{1\over\epsilon}
+{238\over125}
+{12\over25}\gamma_E
+{176\over25}{\zeta^{\prime}(-1)\over\zeta(-1)}
-{38\over25}{\zeta^{\prime}(-3)\over\zeta(-3)}
\right]\left({\mu\over4\pi T}\right)^{6\epsilon}T^4 \, . \nonumber \\
\eqa


\subsubsection{$(hhs)$ contributions}
All the diagrams except $\rm (3f)$ are 
infrared finite in the limit $m_D\rightarrow0$.
This implies that the $g^5$ contribution is given by 
using a dressed longitudinal propagator and bare vertices.
The ring diagram $\rm (3f)$ is infrared divergent in that limit.
The contribution through $g^5$ is obtained by expanding
in powers of self-energies and vertices and one obtains
\bqa\nonumber
{\cal F}^{(hhs)}_{\rm 3a-3g}&=&
-{N_c^2\over4}g^4T(d-1)^4\int_p{1\over(p^2+m_D^2)^2}\sumint_{QR}{1\over Q^2R^2}
\\ && \nonumber
+{N_c^2\over12}g^4T(d-1)^2
\left[d^2-11d+46\right]\int_{p}{p^2\over(p^2+m_D^2)^2}\sumint_{QR}
{1\over Q^2R^4}
\\ \nonumber
&=& 
-{\pi^3\over2}\left({N_c\alpha_s\over3\pi}\right)^2
{T^5\over m_D}
\\ &&-
{33\pi\over16}\left({N_c\alpha_s\over3\pi}\right)^2
\left[
{1\over\epsilon}+{59\over33}+2\gamma_E+2{\zeta^{\prime}(-1)\over\zeta(-1)}
\right]m_DT^3
\left({\mu\over2m_D}\right)^{2\epsilon}\left({\mu\over4\pi T}\right)^{4\epsilon}
\;.
\eqa

\subsubsection{$(hss)$ contribution}
For all the diagrams that are infrared safe, the $(hss)$ contribution is
of order $g^4m_D^2$, i.e. $g^6$ and can be ignored. The infrared divergent diagrams
contribute as follows
\bqa\nonumber
{\cal F}^{(hss)}_{\rm 3a-3g}&=&{1\over4}g^4T^2N_c^2(d-1)^2
\int_{pq}
\left[
{4\over p^2(q^2+m^2_D)(r^2+m_D^2)}
-{2(p^2+4m_D^2)\over p^2(q^2+m^2_D)^2(r^2+m_D^2)}
\right]\sumint_{R}{1\over R^2}\\
&=&
{3\pi^2\over4}
\left[{1\over\epsilon}+1
+2{\zeta^{\prime}(-1)\over\zeta(-1)}\right]
\left({N_c\alpha_s\over3\pi}\right)^2
\left({\mu\over2m_D}\right)^{4\epsilon}\left({\mu\over4\pi T}\right)^{2\epsilon}
T^4\;.
\eqa

\subsubsection{$(sss)$ contribution}
The $(sss)$ contribution to the free energy is given by a three-loop
calculation of the free energy of Electrostatic QCD in three dimensions.
This calculation was performed in Ref.~\cite{BN-96}.
Alternatively,
one can isolate the $(sss)$ contributions from the diagrams listed
in Ref.~\cite{AZ-95}. The result is 
\bqa\nonumber
{\cal F}^{(sss)}_{\rm 3a-3g}&=&
N_c^2g^4T^3\int_{pqr}\left\{
-{1\over4}{1\over(p^2+m^2_D)(q^2+m^2_D)(r^2+m^2_D)^2} 
\right. \\&&\left. \nonumber
+{2\over(p^2+m^2_D)(q^2+m^2_D)(r^2+m^2_D)({\bf q}-{\bf r})^2}
\right. \\&&\left. \nonumber
-{2m_D^2\over(p^2+m^2_D)(q^2+m^2_D)(r^2+m^2_D)^2({\bf q}-{\bf r})^2}
\right. \\&&\left. \nonumber
-{m_D^2\over(p^2+m^2_D)(q^2+m^2_D)(r^2+m^2_D)({\bf q}-{\bf r})^4}
\right. \\&&\left. \nonumber
-{1\over4}
{({\bf p}-{\bf q})^2\over(p^2+m^2_D)(q^2+m^2_D)(r^2+m^2_D)
({\bf q}-{\bf r})^2({\bf r}-{\bf p})^2}
\right. \\&&\left. \nonumber
-{1\over2}(d-2)
{1\over(p^2+m^2_D)(q^2+m^2_D)({\bf q}-{\bf r})^2({\bf r}-{\bf p})^2}
\right. \\&&\left. \nonumber 
+{1\over2}(3-d)
{(r^2+m^2_D)\over(p^2+m^2_D)(q^2+m^2_D)({\bf p}-{\bf q})^2
({\bf q}-{\bf r})^2({\bf r}-{\bf p})^2}
\right. \\&&\left. \nonumber 
-{1\over2}(d-2)
{(r^2+m^2_D)^2\over(p^2+m^2_D)(q^2+m^2_D)({\bf p}-{\bf q})^4
({\bf q}-{\bf r})^2({\bf r}-{\bf p})^2}
\right. \\&&\left. \nonumber 
+{4m_D^2\over(p^2+m^2_D)(q^2+m^2_D)(r^2+m^2_D)({\bf q}-{\bf r})^2({\bf r}-{\bf p})^2}
\right. \\&&\left. \nonumber
+{2m_D^2\over(p^2+m^2_D)(q^2+m^2_D)({\bf p}-{\bf q})^2
({\bf q}-{\bf r})^2({\bf r}-{\bf p})^2}
\right. \\&&\left. \nonumber 
-{4m_D^4\over(p^2+m^2_D)(q^2+m^2_D)(r^2+m^2_D)^2({\bf q}-{\bf r})^2({\bf r}-{\bf p})^2}
\right. \\&&\left. \nonumber
-{3\over8}
{1\over(p^2+m^2_D)(q^2+m^2_D)[({\bf q}-{\bf r})^2+m^2_D][({\bf r}-{\bf p})^2+m^2_D)]}
\right. \\&&\left. \nonumber
-{1\over2}
{({\bf p}-{\bf q})^2
\over(p^2+m^2_D)(q^2+m^2_D)[({\bf q}-{\bf r})^2+m^2_D][({\bf r}-{\bf p})^2+m^2_D]r^2}
\right. \\&&\left. \nonumber
-{1\over4}
{({\bf p}-{\bf q})^4
\over(p^2+m^2_D)(q^2+m^2_D)[({\bf q}-{\bf r})^2+m^2_D][({\bf r}-{\bf p})^2+m^2_D]r^4}
\right. \\&&\left. \nonumber
-{2m_D^2\over(p^2+m^2_D)(q^2+m^2_D)[({\bf q}-{\bf r})^2+m^2_D][({\bf r}-{\bf p})^2+m^2_D)]r^2}
\right. \\&&\left. \nonumber
-m_D^2
{({\bf p}-{\bf q})^2
\over(p^2+m^2_D)(q^2+m^2_D)[({\bf q}-{\bf r})^2+m^2_D][({\bf r}-{\bf p})^2+m^2_D]r^4}
\right. \\&&\left. \nonumber
-m_D^4
{1\over(p^2+m^2_D)(q^2+m^2_D)
[({\bf q}-{\bf r})^2+m^2_D][({\bf r}-{\bf p})^2+m^2_D]r^2({\bf p}-{\bf q})^2}
\right. \\&&\left. \nonumber
-{m_D^4\over(p^2+m^2_D)(q^2+m^2_D)
[({\bf q}-{\bf r})^2+m^2_D][({\bf r}-{\bf p})^2+m^2_D]r^4}
\right. \\&&\left. 
-{1\over4}{(q^2+m^2_D)\over(p^2+m^2_D)[({\bf r}-{\bf p})^2+m^2_D]
[({\bf q}-{\bf r})^2+m^2_D]r^2({\bf p}-{\bf q})^2}
\right\}\;.
\eqa

The expression for the integrals are given in Appendix C. Adding
Eqs.~(\ref{sssfirst})--~(\ref{ssslast}), the final result
is
\bqa
{\cal F}^{(sss)}_{\rm 3a-3g}&=&{9\pi\over4}\left({N_c\alpha_s\over3\pi}\right)^2
\left[{89\over24}-{11\over6}\log2
+{1\over6}\pi^2\right]m_DT^3
\;.
\eqa
Note that all the poles in $\epsilon$ cancel.

\section{The thermodynamic potential}

In this section we present the final renormalized 
thermodynamic potential explicitly through order
$\delta^2$, aka NNLO.  The final NNLO expression
is completely analytic; however, there are some numerically
determined constants which remain in the final expressions at
NLO.

\subsection{Leading order}
The leading order thermodynamic potential
is given by the contribution from the diagrams $\rm (1a)$ and $\rm (1b)$.

\bqa
\Omega_{\rm 1-loop} &=& 
{\cal F}_{\rm ideal}
\left\{ 1 - {15 \over 2} \hat m_D^2 
+ 30 \hat m_D^3
+ {45 \over 8}
\left({1\over\epsilon} +2\log {\hat \mu \over 2}
        - 7 + 2\gamma_E + {2\pi^2\over 3} \right)
        \hat m_D^4\right\} , \nonumber \\ 
\label{Omega-1}
\eqa
where ${\cal F}_{\rm ideal}$ is the free energy of a gas of $N_c^2-1$
massless spin-one bosons and
$\hat m_D$ and $\hat \mu$ are dimensionless variables:
\bqa
{\cal F}_{\rm ideal}&=&\left(N_c^2-1\right)\left(-{\pi^2\over45}T^4\right)\;,\\
\hat m_D &=& {m_D \over 2 \pi T}  \;,
\\
\hat \mu &=& {\mu \over 2 \pi T}  \;. 
\eqa
%
The complete expression for the leading order thermodynamic potential
is given by 
adding the leading vacuum energy counterterm~(\ref{lovac}) to 
Eq.~(\ref{Omega-1}):
\bqa
\Omega_{\rm LO} &=& 
{\cal F}_{\rm ideal}
\left\{ 1 - {15 \over 2} \hat m_D^2 
+ 30 \hat m_D^3
+ {45 \over 4}
\left(\log {\hat \mu \over 2}
        - {7\over2} + \gamma_E + {\pi^2\over 3} \right)
        \hat m_D^4\right\}\;.
\eqa

\subsection{Next-to-leading order}
The renormalization contributions at first order in $\delta$ are
\bqa
\Delta\Omega_1&=&\Delta_1{\cal E}
+\Delta_1m_D^2{\partial\over\partial m_D^2}\Omega_{\rm LO}
\eqa
%
Using the results listed in Eqs.~(\ref{del111}) 
and (\ref{del333}) the complete contribution from the counterterm at 
first order in $\delta$ is
\bqa
\nonumber 
\Delta\Omega_1&=& 
{\cal F}_{\rm ideal}
\Bigg\{ {45\over4\epsilon} \hat m_D^4 
	+{165\over8}
\Bigg[
{1\over\epsilon}+2\log{\hat\mu\over 2} + 2 {\zeta'(-1)\over\zeta(-1)}+2 \bigg]{N_c\alpha_s \over 3\pi}  
\hat m_D^2
\\ && 
-{495\over4}
		\left[{1\over\epsilon}+2\log{\hat\mu\over 2} - 2 \log \hat m_D +2 \right] {N_c\alpha_s \over 3\pi}  \hat m_D^3\bigg\}
\;. 
\label{OmegaVMct1}
\eqa
The NLO thermodynamic potential reads
\bqa
\Omega_{\rm NLO}&=&
{\cal F}_{\rm ideal}\Bigg\{ 
	1  - 15 \hat m_D^3 
	- {45\over4}\left(\log\hat{\mu\over2}-{7\over2}+\gamma_E+{\pi^2\over3}
\right)\hat m_D^4+
\Bigg[ -{15\over4}
	+ 45 \hat m_D
\nonumber \\ && 
\nonumber
	-{165\over4}
\left(\log{\hat\mu \over 2}-{36\over11}\log\hat{m}_D
-2.001\right)\hat m_D^2
	+{495\over2}\left(\log{\hat\mu \over 2}+{5\over22}
+\gamma_E\right)\!\!\hat m_D^3
\Bigg]
	{N_c\alpha_s\over3\pi} 
\Bigg\} \;.
\\ && 
\label{Omega-NLO}
\eqa
%
\subsection{Next-to-next-to-leading order}

The renormalization contributions at second order in $\delta$ are
\bqa\nonumber
\Delta\Omega_2&=&\Delta_2{\cal E}_0
+\Delta_2m_D^2{\partial\over\partial m_D^2}\Omega_{\rm LO}
+\Delta_1m_D^2{\partial\over\partial m_D^2}\Omega_{\rm NLO}
+{1\over2}\left({\partial^2\over(\partial m_D^2)^2}
\Omega_{\rm LO}\right)\left(\Delta_1m_D^2\right)^2
\\ &&
+(N_c^2-1){{\cal F}_{\rm 2a+2b+2c}\over\alpha_s}\Delta_1\alpha_s
\;.
\eqa

Using the results listed above, we obtain
\bqa\nonumber
\Delta\Omega_2&=& 
{\cal F}_{\rm ideal}
\Bigg\{ 
-{45\over8\epsilon}\hat{m}_D^4			
-{165\over8}
{N_c\alpha_s \over 3\pi}  
\Bigg[
{1\over\epsilon}+2\log{\hat\mu\over 2} + 2 {\zeta'(-1)\over\zeta(-1)}+2 \bigg]
\hat m_D^2
\\ && \nonumber
+{1485\over8}{N_c\alpha_s \over 3\pi}\left[
{1\over\epsilon}
+2\log{\hat\mu\over 2} -2\log{\hat{m}_D}+{4\over3}
\right]\hat{m}_D^3
\\ &&\nonumber
+\left({N_c\alpha_s\over3\pi}\right)^2\left[
{165\over16}
\left(
{1\over\epsilon}+4\log{\hat{\mu}\over2}
+2+4{\zeta^{\prime}(-1)\over\zeta(-1)}\right)
\right.\\ &&\left.
-{1485\over8}\left({1\over\epsilon}+4\log{\hat{\mu}\over2}
-2\log\hat{m}_D
+{4\over3}+2{\zeta^{\prime}(-1)\over\zeta(-1)}
\right)\hat{m}_D
\right]
\Bigg\}\;. 
\label{OmegaVMct2}
\eqa

Adding the NNLO counterterms (\ref{OmegaVMct2}) to the contributions from the 
various NNLO diagrams we obtain the
renormalized NNLO thermodynamic potential.  We note that at NNLO all
numerically determined coefficients of order $\epsilon^0$ drop out and 
we are left with a final result which
is completely analytic.
The resulting NNLO thermodynamic potential is

\begin{eqnarray}\nonumber
\Omega_{\rm NNLO}&=&
{\cal F}_{\rm ideal}
\bigg\{1-{15\over4}\hat{m}_D^3
+{N_c\alpha_s\over3\pi}\left[
-{15\over4}+{45\over2}\hat{m}_D-{135\over2}\hat{m}^2_D
-{495\over4}\left(\log\hat{\mu\over2}
+{5\over22}+\gamma_E
\right)\hat{m}^3_D
\right]
\\ && \nonumber
+\left({N_c\alpha_s\over3\pi}\right)^2\left[{45\over4\hat{m}_D}
-{165\over8}\left(\log{\hat{\mu}\over2}
-{72\over11}\log{\hat{m}_D}
-{84\over55}
-{6\over11}\gamma_E
-{74\over11}{\zeta^{\prime}(-1)\over\zeta(-1)}
\right.\right.\\&&\left.\left.
+{19\over11}{\zeta^{\prime}(-3)\over\zeta(-3)}
\right)
+{1485\over4}
\left(\log{\hat{\mu}\over2}
-{79\over44}+
\gamma_E
+\log2-{\pi^2\over11}
\right)\hat{m}_D
\right]
\bigg\}\;.
\label{Omega-NNLO}
\end{eqnarray}

%
Note that if we use the weak-coupling value for the Debye
mass $m_D^2=4\pi N_c\alpha_sT^2/3$, the NNLO HTLpt result~(\ref{Omega-NNLO})
is guaranteed to
reduce to the weak-coupling result through order $g^5$ and we have checked
that this is the case.

%

%
%
\section{Thermodynamic functions}
\subsection{Mass prescriptions}
The mass parameter $m_D$ in HTLpt
is completely arbitrary. To complete a 
calculation, it is necessary to specify $m_D$ as
function of $g$ and $T$.  In this section we will discuss
several prescriptions for the mass parameter.

\subsubsection{Variational Debye mass}
The variational mass is given by the solution to 
the equation
\bqa
{\partial \ \ \over \partial m_D}\Omega(T,\alpha_s,m_D,\mu,\delta=1) = 0\;.
\eqa 
This yields
\begin{eqnarray}
{45\over4} \hat m_D^2 &=& {N_c\alpha_s\over3\pi} \left[{45\over2} - 135 \hat m_D - {1485\over4} \left( \log{\hat\mu\over2} + {5\over22} + \gamma_E\right)\!\!\hat m_D^2 \right]
\nonumber \\ &&
+ \left({N_c\alpha\over3\pi}\right)^2 \left[-{45\over4}{1\over\hat m_D^2} + {135\over \hat m_D} + {1485\over4}\left(\log{\hat\mu \over 2} - {79\over44} + 
\gamma_E + \log2 - {\pi^2\over11}\right)\right]\;.
\label{gapleik}
\end{eqnarray}

At leading order in HTLpt, the only solution is the trivial solution, i.e.
$m_D=0$. In that case, it is natural to chose the weak-coupling result for
the Debye mass. This was done in Ref.~\cite{htlpt1}.

At NLO, the resulting gap equation has a nontrivial
solution, which is real for all values of the coupling.
At NNLO, the solution to the gap equation~(\ref{gapleik}) is plagued
by imaginary parts for all values of the coupling. The problem with
complex solutions
seems to be generic
since it has also been observed in screened 
perturbation
theory~\cite{spt} and QED~\cite{qednnlo,qednnlo1,qednnlo2}.
In those cases, however, it was complex only for small values of the coupling.

\subsubsection{Perturbative Debye mass}
At leading order in the coupling constant $g$, the Debye mass is given by
the static longitudinal gluon self-energy at zero three-momentum, 
$m_D^2=\Pi_L(0,0)$, i.e.
\bqa\nonumber
m_D^2&=&N_c(d-1)^2g^2\sumint_P{1\over P^2}\\
&=&{4\pi\over3}N_c\alpha_sT^2\;.
\eqa
The next-to-leading order correction to the Debye mass is determined
by resummation of one-loop diagrams with dressed vertices. 
Furthermore, since it suffices to
take into account static modes in the loops, the HTL-corrections to
the vertices also vanish. The result, however, turns out to
be logarithmically infrared divergent, which reflects the sensitivity
to the nonperturbative magnetic mass scale. 
The result was first obtained by Rebhan~\cite{antonqcd}
and 
reads~
\bqa
\delta m^2_D&=&m_D^2\sqrt{3N_c\over\pi}\alpha^{1/2}
\left[\log{2m_D\over m_{\rm mag}}-{1\over2}
\right]\;,
\eqa
where $m_{\rm mag}$ is the nonperturbative magnetic mass.
We will not use this mass prescription since it involves
the magnetic mass which would require input from e.g. lattice simulations.

\subsubsection{BN mass parameter $m_E^2$}
In the previous subsection, we saw that the Debye mass is sensitive to
the nonperturbative magnetic mass which is of order $g^2T$.
In QED, the situation is much better. The Debye mass can be calculated
order by order in $e$ using resummed perturbation theory.
The Debye mass then receives contributions from the scale $T$
and $eT$.
Effective field theory methods and dimensional reduction
can be conveniently used to 
calculate separately the contributions to $m_D$ 
from the two scales in the problem.
The contributions to $m_D$ and other physical quantities
from the scale $T$ can be calculated
using bare propagators and vertices.
The contributions from the soft scale can be calculated using an effective
three-dimensional field theory called electrostatic QED.
The parameters of this effective theory are obtained by a matching 
procedure and encode the physics from the scale $T$.
The effective field theory contains a massive
field $A_0$ that up to normalization
can be identified with the zeroth component of the gauge field in QED.
The mass parameter $m_E$ of $A_0$ gives the
contribution to the Debye mass from the hard scale $T$ and can
be written as a power series in $e^2$. 

For nonabelian gauge theories, the corresponding effective three-dimensional
theory is called electrostatic QCD. The mass parameter $m_E$
for the field $A_0^a$ (which lives in the adjoint representation) 
can also be calculated as a power series in $g^2$.
It has
been determined to order $g^4$ by Braaten and Nieto~\cite{BN-96}.
For pure-glue QCD, it reads
\bqa
m_E^2&=&
{4\pi\over3}N_c\alpha_sT^2\left[
1+{N_c\alpha_s\over3\pi}\left(
{5\over4}+{11\over2}\gamma_E
+{11\over2}\log{\mu\over4\pi T}
\right)
\right]\;.
\eqa
We will use the mass parameter $m_E$ as another prescription for the Debye
mass and denote it by the Braaten Nieto (BN) mass prescription.

\subsection{Pressure}
In this subsection, we present our results for the pressure using the
variational mass prescription and the BN mass prescription.

\subsubsection{Variational mass}

\FIGURE{\includegraphics[width=9.55cm]{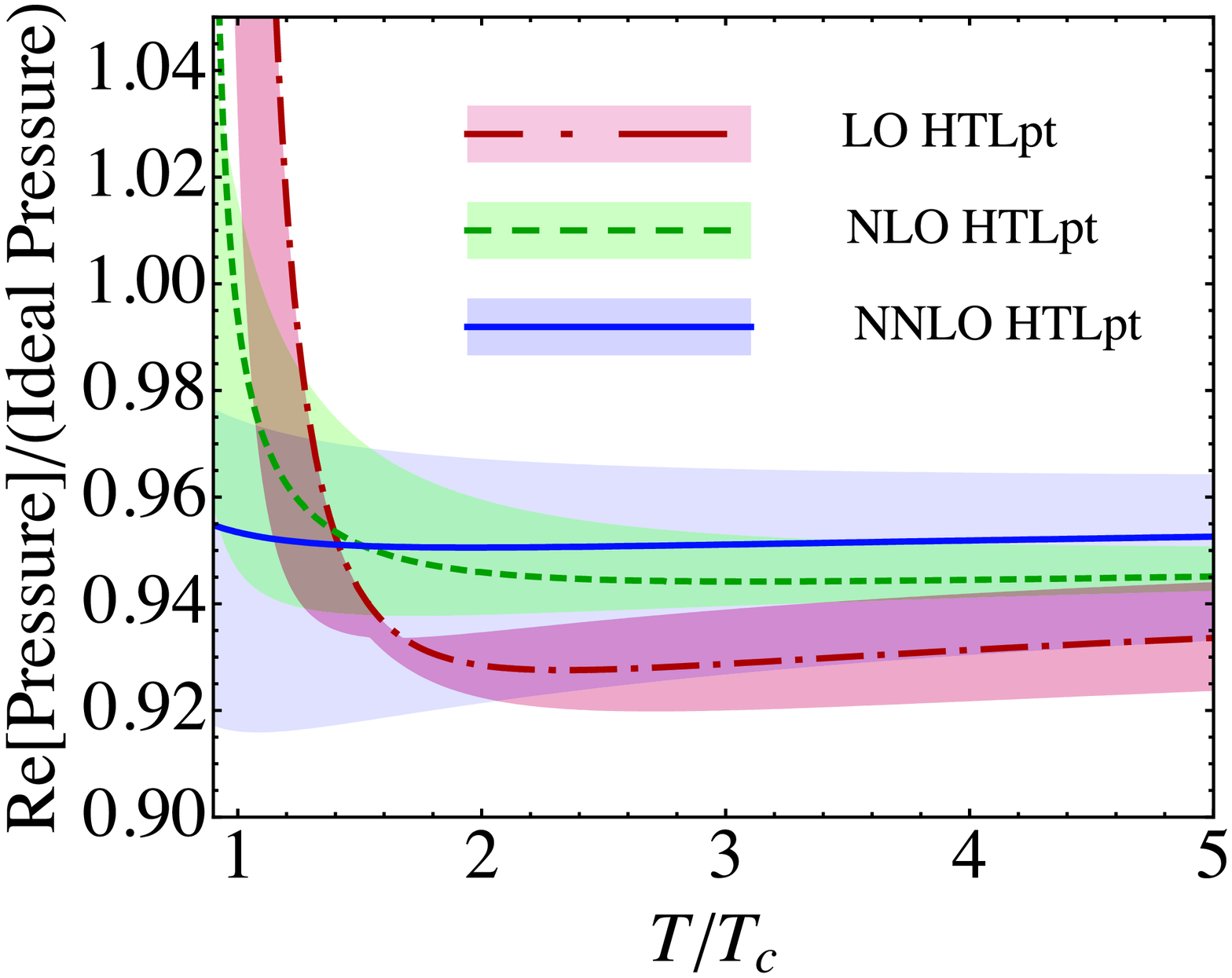}
\caption{Comparison of LO, NLO, and NNLO predictions for the 
scaled real part of the pressure using the variational mass prescription. 
Shaded
bands show the result of varying the renormalization scale $\mu$ by a factor 
of two around $\mu = 2 \pi T$.}
\label{fig:pressure}}

In Fig.~\ref{fig:pressure}, we compare the LO, NLO, and NNLO predictions for
the real part of the pressure normalized to that of an ideal gas.
 Shaded
bands show the result of varying the renormalization scale $\mu$ by a factor 
of two around $\mu = 2 \pi T$. We use three-loop running of $\alpha_s$.

\FIGURE{\includegraphics[width=9.55cm]{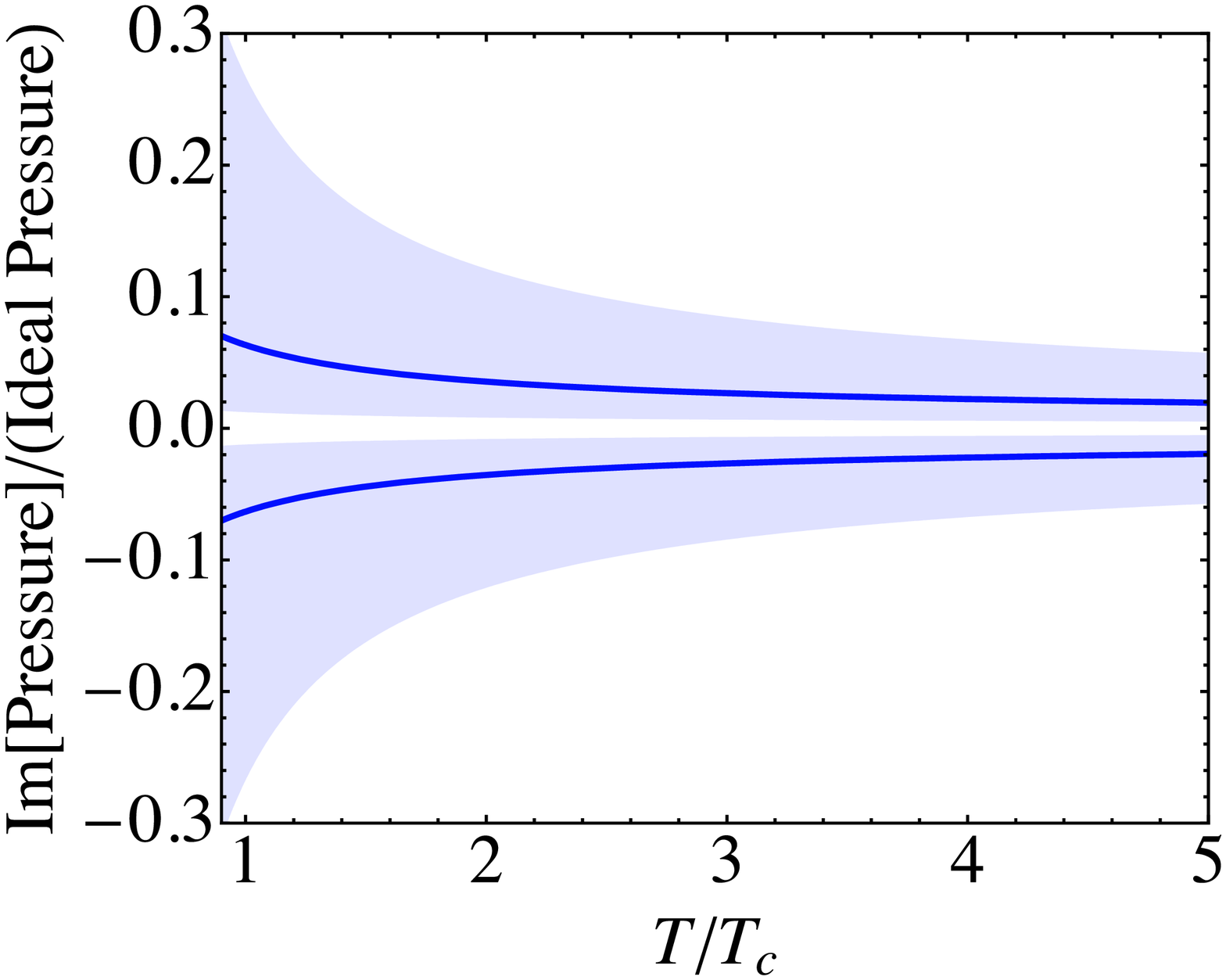}
\caption{The NNLO result for the scaled 
imaginary part of the pressure with three-loop running
and variational mass prescription. 
The two curves arise from the two
complex conjugate solutions to the gap equations.
Shaded
bands show the result of varying the renormalization scale $\mu$ by a 
factor of two
around $\mu = 2 \pi T$.}
\label{imagp}}

In Fig.~\ref{imagp}, we show the NNLO result for the imaginary part of the
pressure normalized by the ideal gas pressure.
We use three-loop running of $\alpha_s$.
The imaginary part decreases with increasing temperature and is rather
small beyond 3-4 $T_c$.

\subsubsection{BN mass}

\FIGURE{\includegraphics[width=9.55cm]{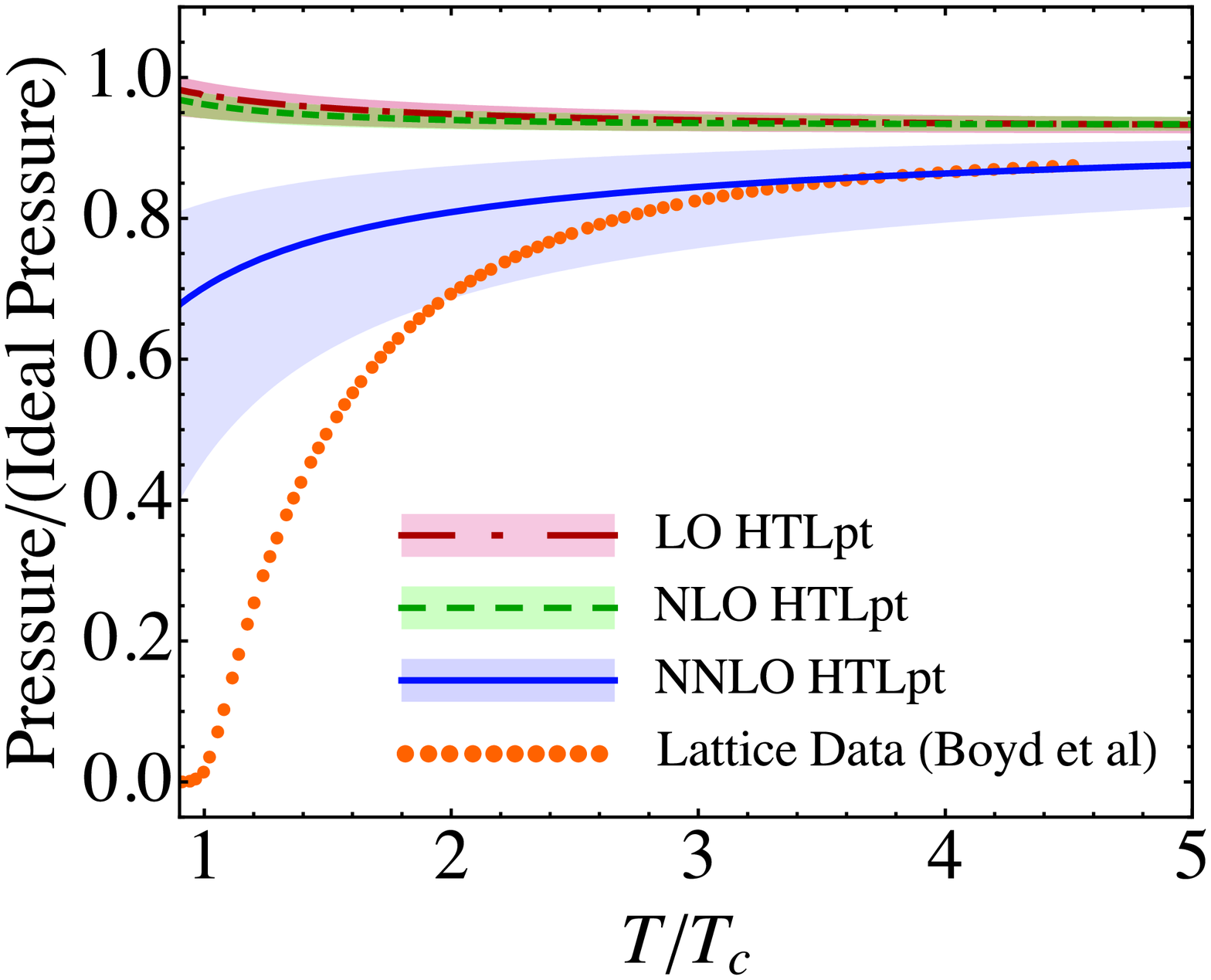}
\caption{Comparison of LO, NLO, and NNLO predictions for the scaled pressure 
using the BN mass prescription and one-loop running of $\alpha_s$.
The points are
lattice data for pure-glue with $N_c=3$
from Boyd et al. \cite{Boyd:1996bx}.  
Shaded bands show the result of varying the 
renormalization scale $\mu$ by a factor of two
around $\mu = 2 \pi T$.}
\label{bn1pressure}}

In Fig.~\ref{bn1pressure}, we show the HTLpt predictions for the pressure
normalized to that of an ideal gas using the BN mass prescription.
The bands are obtained by varying the renormalization scale by a factor of
two
around $\mu=2\pi T$. We use one-loop running of $\alpha_s$.
In Fig.~\ref{bn3pressure}, we again plot the normalized pressure, but now with
three-loop running of $\alpha_s$. The agreement between the lattice data
from Boyd et al. \cite{Boyd:1996bx} is very good down to temperatures of
around 3 $T_c$. Comparing Figs.~\ref{bn1pressure}--\ref{bn3pressure}
we see that using the three-loop running, the band becomes wider.
However, the difference is significant only for low $T$, where the HTLpt results
disagrees with the lattice anyway. For $T>3T_c$, the prescription for the
running makes very little difference.

\FIGURE{\includegraphics[width=9.55cm]{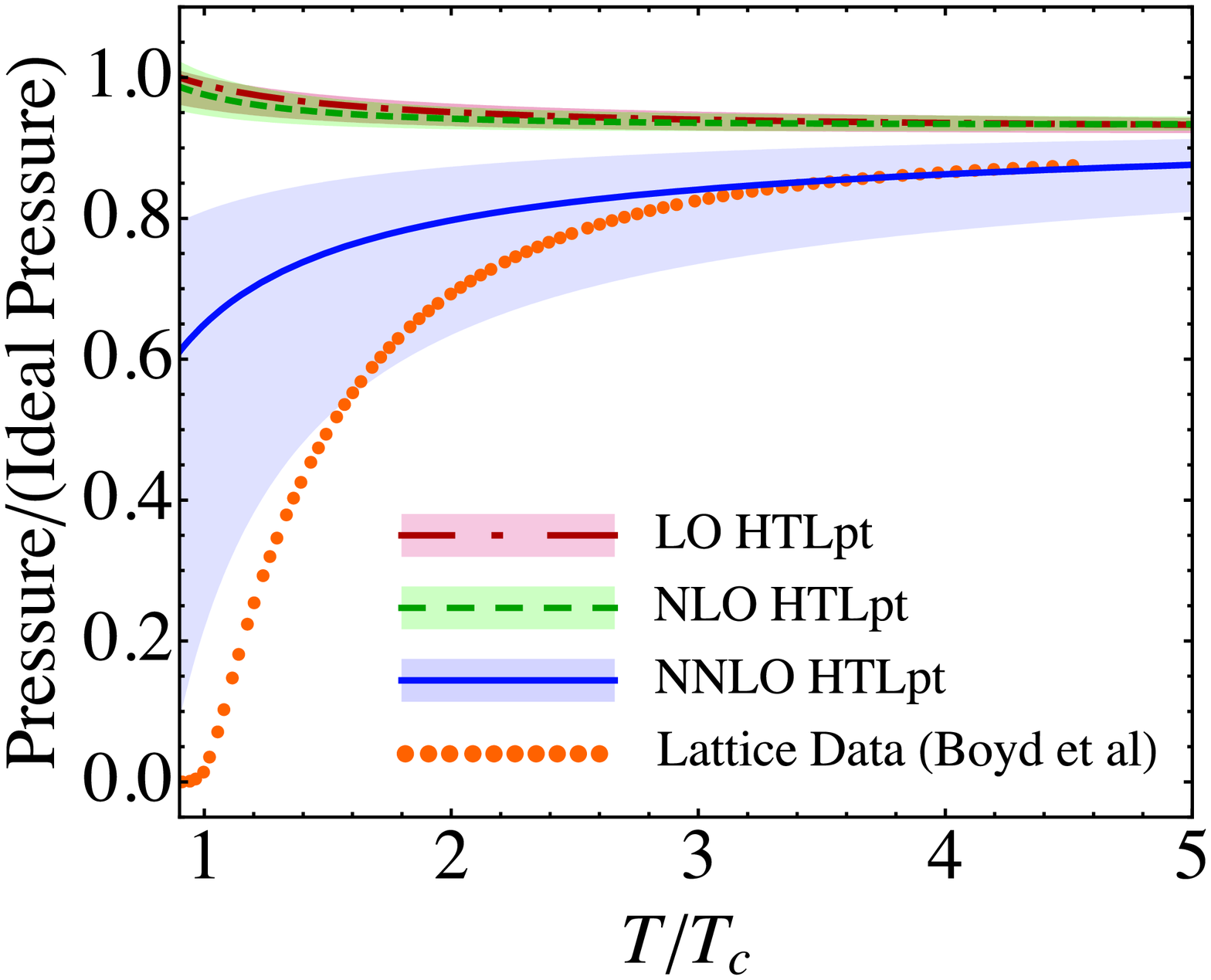}
\caption{Comparison of LO, NLO, and NNLO predictions for the scaled pressure 
using the BN mass prescription and three-loop running of $\alpha_s$.
The points are lattice data
pure-glue with $N_c=3$ from Boyd et al. \cite{Boyd:1996bx}.  Shaded
bands show the result of varying the renormalization scale $\mu$ by a factor of two
around $\mu = 2 \pi T$.}\label{bn3pressure}}

Until recently, lattice data for thermodynamic variables only existed for
temperatures up to approximately 5 $T_c$. In the paper
by Enrodi et al~\cite{endrodi}, the authors
calculate the pressure on the lattice for pure-glue QCD at very large
temperatures. In Fig.~\ref{highpressure}, we show the results of Enrodi
et al as well as Boyd et al, together with 
the HTLpt NLO and NNLO predictions for the pressure.
The two points from Ref.~\cite{endrodi} have large error bars, but
the data points are consistent with the HTL predictions.

\FIGURE{\includegraphics[width=9.55cm]{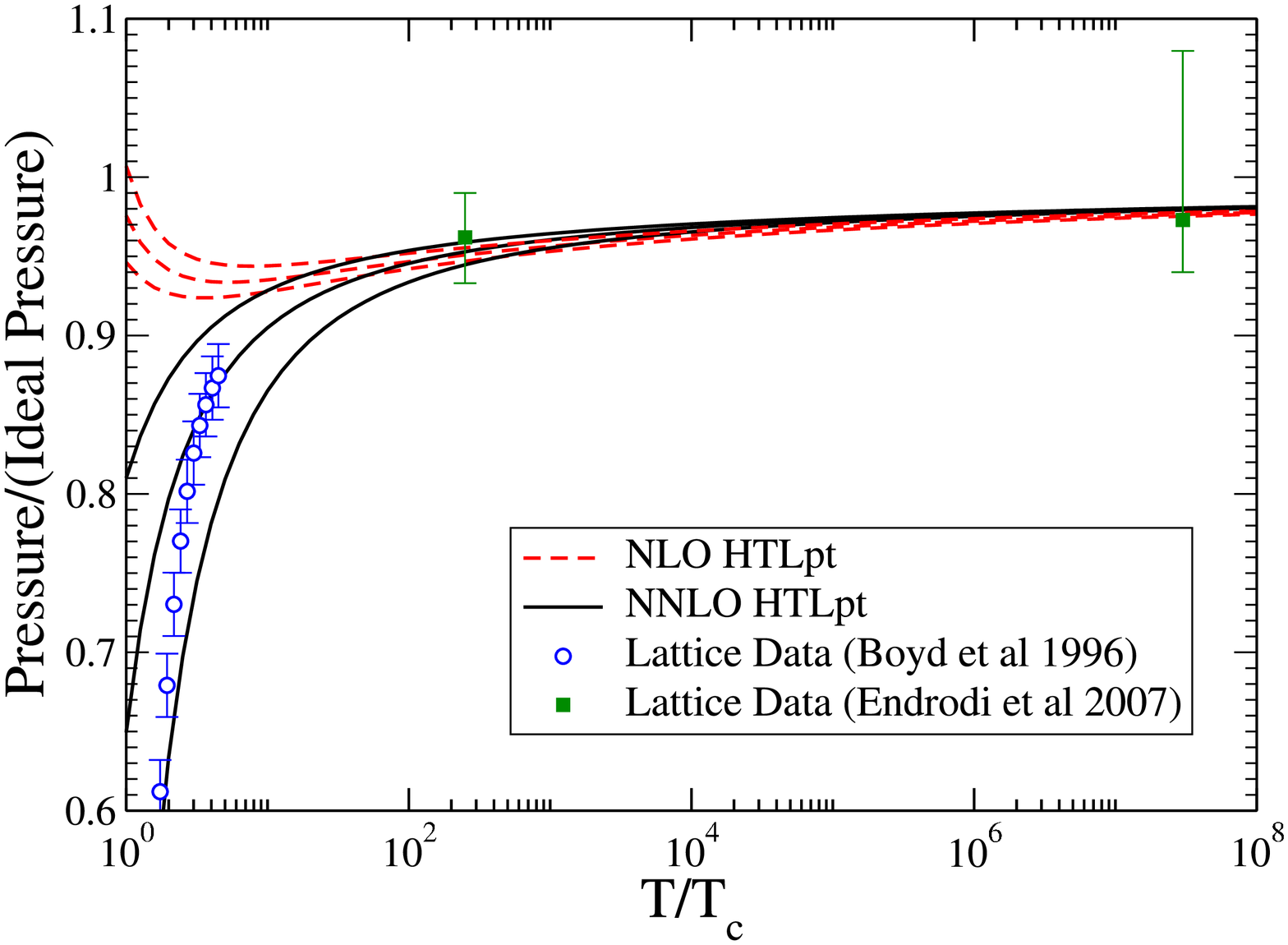}
\caption{Comparison of NLO, and NNLO predictions for the scaled pressure 
with SU(3) pure-glue lattice data from Boyd et al. \cite{Boyd:1996bx}
and Endrodi et al.~\cite{endrodi}.
Shaded
bands show the result of varying the renormalization scale $\mu$ by a factor 
of two
around $\mu = 2 \pi T$.}
\label{highpressure}}


We note that our prediction for the normalized free energy using either
the
leading-order, BN, or variational mass prescriptions is independent of
$N_c$
if one holds $\alpha_s N_c$ fixed ('t Hooft limit).  However, this need
not
be the case for an arbitrary mass prescription.  The $N_c$-independence of
the normalized free energy of the free energy is in agreement with
recent lattice measurements that show a very small dependence of the free
energy on $N_c$~\cite{Panero:2009tv,Datta:2010sq}.

Due to the imaginary parts we do not show results coming from the 
variational prescription in the remainder of the results section.  We will
return to a discussion of the dependence on mass prescriptions in 
the conclusions.

\subsection{Energy density}
The energy density $\cal E$ is defined by
\bqa
{\cal E}&=&{\cal F}-T{d{\cal F}\over dT}\;.
\eqa
In Fig.~\ref{fig:energy}, we show the 
LO, NLO, and NNLO predictions for energy density normalized to that of
an ideal gas.
We use three-loop running and the BN mass.
The bands show the result of varying the renormalization scale $\mu$ by a 
factor of two
around $\mu = 2 \pi T$.
Our NNLO predictions are in excellent agreement with the lattice data
down to $T\simeq2T_c$.

\FIGURE{\includegraphics[width=9.55cm]{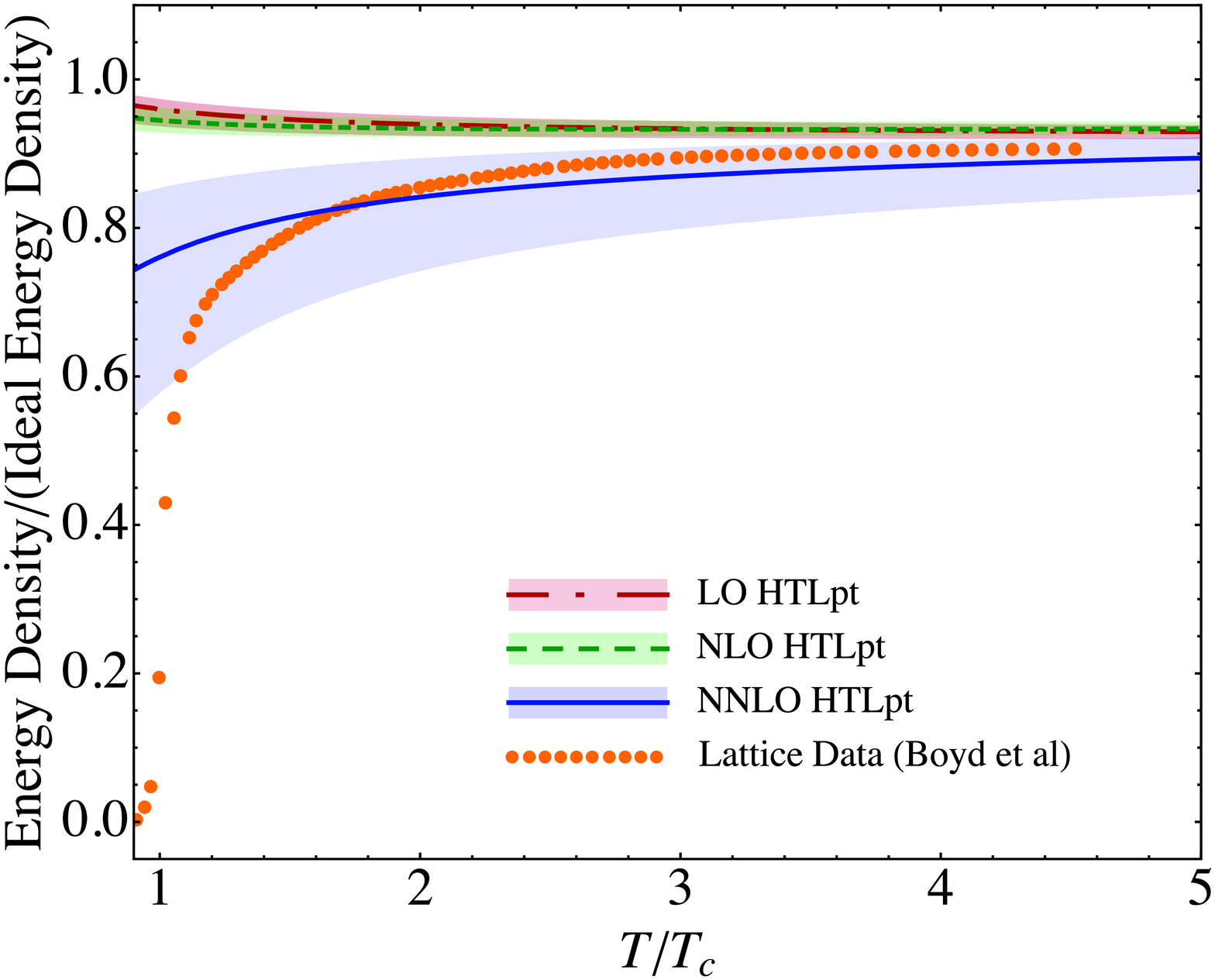}
\caption{Comparison of LO, NLO, and NNLO predictions for the scaled energy 
density 
with SU(3) pure-glue lattice data from Boyd et al. \cite{Boyd:1996bx}.  
We use three-loop running and the BN mass.
Shaded
bands show the result of varying the renormalization scale $\mu$ by a 
factor of two
around $\mu = 2 \pi T$.}
\label{fig:energy}}


\subsection{Entropy}
The entropy density is defined by
\bqa
{\cal S}&=&-{\partial{\cal F}\over\partial T}\;,
\eqa
where all other parameters that ${\cal F}$ depends on, are kept fixed.
In Fig.~\ref{fig:entropy}, we show the entropy density normalized to that
of an ideal gas. We use three-loop running and the BN mass.
The points are lattice data from 
Boyd et al. \cite{Boyd:1996bx}.  
Our NNLO predictions are in excellent agreement with the lattice data
down to $T\simeq2T_c$.

\FIGURE{\includegraphics[width=9.55cm]{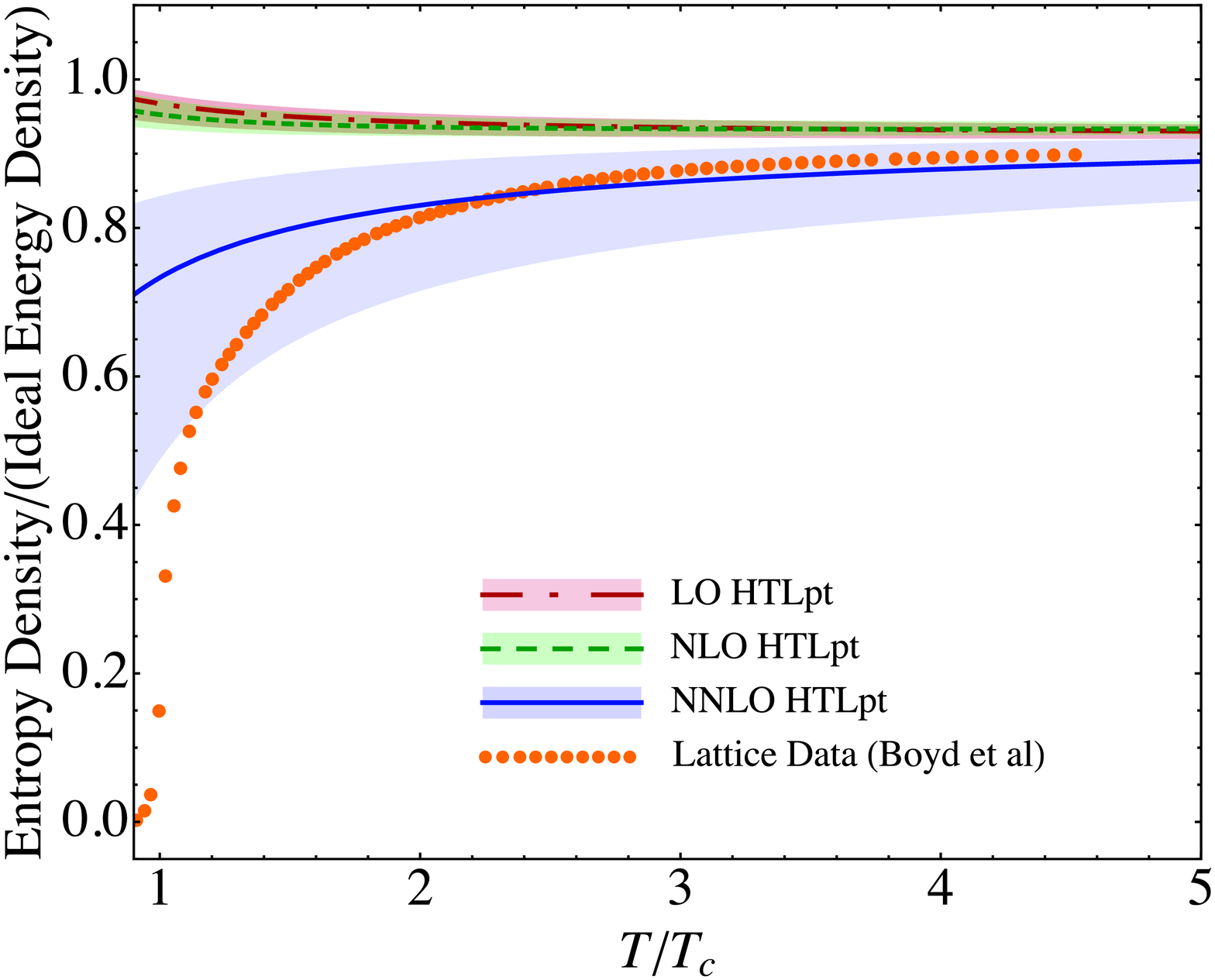}
\caption{Comparison of LO, NLO, and NNLO predictions for the scaled entropy 
with SU(3) pure-glue lattice data from Boyd et al. \cite{Boyd:1996bx}.  
We use three-loop running and the BN mass.
Shaded
bands show the result of varying the renormalization scale $\mu$ by a factor 
of two
around $\mu = 2 \pi T$.}
\label{fig:entropy}}


\subsection{Trace anomaly}
In pure-glue QCD or in QCD with massless quarks, there is no mass scale
in the Lagrangian and the theory is scale invariant.
At the classical level, this implies that the
trace of the energy-momentum tensor vanishes.
At the quantum level, scale invariance is broken by renormalization effects.
It is convenient to introduce the scale anomaly density
${\cal E}-3{\cal P}$, which is proportional to the 
trace of the energy-momentum tensor. The trace anomaly can be written as
\bqa
{\cal E}-3{\cal P}
&=&-T^5{d\over dT}\left({{\cal F}\over T^4}\right)\;.
\eqa
In Fig.~\ref{trace}, we show the HTLpt predictions for the trace anomaly 
divided by
${\cal E}_{\rm ideal}$ using the BN mass prescription and three-loop running of $\alpha_s$.
The points are lattice data from Boyd et al. \cite{Boyd:1996bx} 
For temperatures below approximately 2$T_c$, there is a large discrepancy
between the HTLpt predictions and lattice data. At LO and NLO, the 
curves are even bending downwards.

At temperatures close to the phase transition it has been suggested that the 
discrepancy between HTLpt resummed predictions for thermodynamics 
functions and, in particular, the trace anomaly is due to influence of a 
power corrections~\cite{Pisarski:2000eq,Kondo:2001nq,Pisarski:2002ji,Pisarski:2006yk,Narison:2009ag}
which are related to confinement.  Phenomenological fits of lattice data which 
include such power corrections show that the agreement with lattice data is improved 
\cite{Megias:2009mp,Megias:2009ar}.  Alternatively, others have constructed
AdS/CFT inspired models which break conformal invariance 
``by hand''~\cite{Andreev:2007zv,Gubser:2008ny,Gubser:2008yx,Noronha:2009ud}.
These models are also able to fit the thermodynamical functions of QCD at temperatures
close to the phase transition.

\FIGURE{\includegraphics[width=9.55cm]{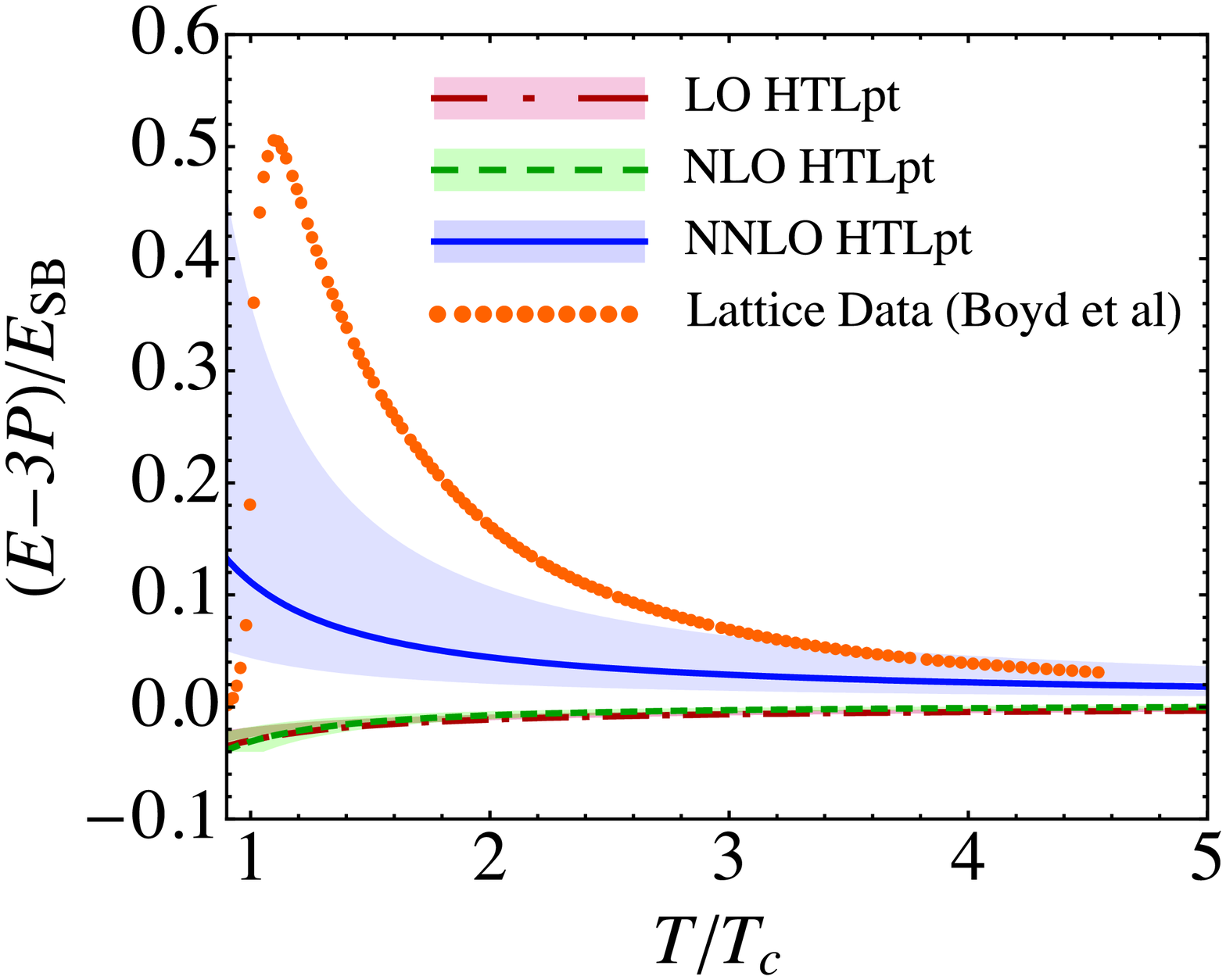}
\caption{Comparison of LO, NLO, and NNLO predictions for the scaled trace
anomaly
with SU(3) pure-glue lattice data from Boyd et al. \cite{Boyd:1996bx}.  
Shaded bands show the result of varying the renormalization scale $\mu$ by a 
factor of two
around $\mu = 2 \pi T$.}
\label{trace}}

In Fig.~\ref{trace2}, we show 
the HTLpt predictions for the trace anomaly scaled by $T^2/T_c^2$
using the BN mass prescription and three-loop running of $\alpha_s$.
The points are lattice data from Boyd et al. \cite{Boyd:1996bx}.
The most remarkable feature is that lattice data are essentially constant
over a very large temperature range.
Clearly, HTLpt does not reproduce the scaled lattice data precisely;
however, the agreement is dramatically improved when going from NLO to NNLO.

\FIGURE{
\includegraphics[width=9.55cm]{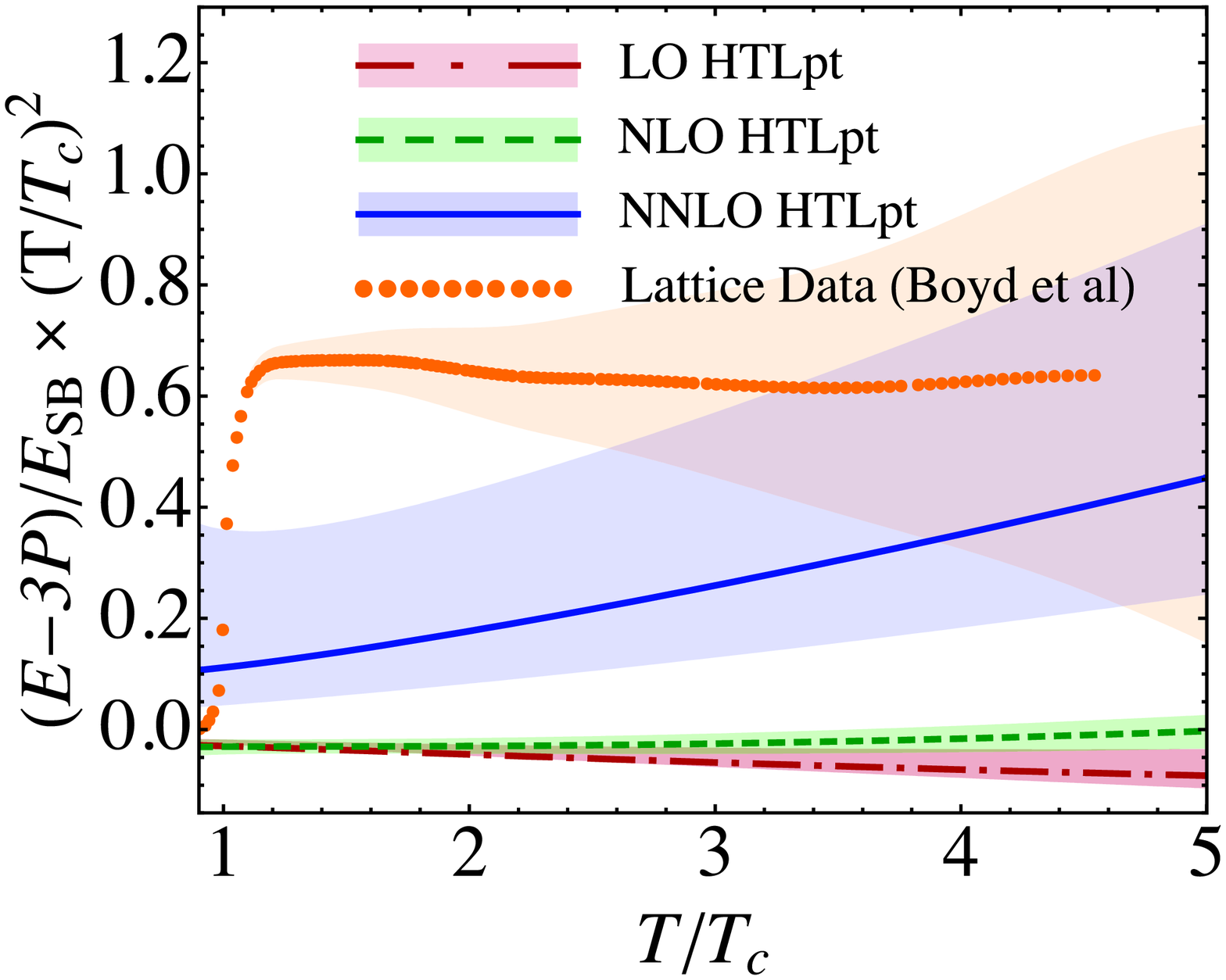}
\caption{Comparison of LO, NLO, and NNLO predictions for the scaled trace
anomaly multiplied by $T^2/T_c^2$
with SU(3) pure-glue lattice data from Boyd et al. \cite{Boyd:1996bx}.  
Shaded bands show the result of varying the renormalization scale $\mu$ by a 
factor of two
around $\mu = 2 \pi T$. Orange band shown with lattice data indicates scaled error assuming
a $\pm2.5\%$ error in the lattice data for the trace anomaly.}
\label{trace2}}

\section{Conclusions}

In this paper, we have presented results for the LO, NLO, and NNLO 
thermodynamic
functions for SU($N_c$) Yang-Mills theory using HTLpt.
We 
compared our predictions 
with lattice data for $N_c=3$ and found that with a perturbative 
mass prescription HTLpt is consistent with 
available lattice data down to approximately $T \sim 3\,T_c$ in the case of the pressure and 
$T \sim 2\,T_c$ in the case 
of the energy density and entropy.  These results are in line with 
expectations since
below $T \sim 2-3\,T_c$ a simple ``electric'' quasiparticle 
approximation breaks down due to nonperturbative chromomagnetic effects.  

The mass parameter $m_D$ in HTLpt is arbitrary and 
we employed two different
prescriptions for fixing it.
Unfortunately, the variational gap equation
has four complex conjugate solutions, two with positive real parts. 
This has also been observed
in scalar theory and QED. Whether this is a problem of HTLpt as such or
is related to our $m_D/T$ expansion is unknown.
Since it is not currently possible to evaluate the NNLO HTLpt
diagrams in gauge theories exactly, it is impossible to settle the issue at
this stage.
On the other hand, the BN mass prescription
is well defined to all orders in perturbation theory and 
does a reasonable job reproducing available
lattice data for temperatures above $T\simeq3T_c$.

That being said without lattice data to compare with
one would be hard pressed to favor one prescription over the other.
While it is true that variational solutions are complex, one could be tempted to 
ignore the imaginary contributions and take ${\rm Re}[{\cal P}]$
as the approximation.  As we have shown (see Fig.~\ref{fig:pressure}) in
this case the convergence of HTLpt is impressive; however, the result lies
above the lattice data at low temperatures.  Taking the BN mass prescription
on the other hand results in a real-valued pressure which seems to be
in better agreement with the lattice data at the expense of poorer convergence
of the successive approximations.  
The dependence on the mass prescription
gives another measure of our theoretical uncertainty.  If one were able to
carry the HTLpt $\delta$ expansion to higher and higher orders the dependence
on the mass prescription would go away; however, going to higher orders
presents a rather daunting task since one encounters a purely non-perturbative
contribution at ${\cal O}(\alpha_s^3)$ for which lattice input is required.

Finally, we emphasize that HTLpt is gauge invariant by construction and
that it can be used to calculate time-dependent quantities as well
since it is formulated in Minkowski space.
The fact that our BN mass results are consistent with lattice data down to 
approximately $3T_c$, suggests that HTLpt can provide a coherent framework
that can be used to systematically calculate 
real-time quantities such as heavy quark diffusion and viscosity 
at temperatures relevant for LHC.

\acknowledgments
We thank H. St\"ocker for his encouragement and support for this endeavor.
N. Su thanks M. Huang, Q. Wang, and
the Department of Physics at the Norwegian University of Science and
Technology for hospitality. 
N. Su was supported by the Frankfurt
International Graduate School for Science and the Helmholtz Graduate School
for Hadron and Ion Research.
M. Strickland was supported in part 
by the Helmholtz International Center for FAIR Landesoffensive zur Entwicklung 
Wissenschaftlich-\"Okonomischer Exzellenz program.

\appendix

\section{HTL Feynman rules}
\setcounter{equation}{0}
\label{app:rules}
In this appendix, we present Feynman rules for HTL perturbation theory 
in QCD.  We give explicit expressions for the propagators 
and for the 3- and 4-gluon  vertices.
The Feynman rules are given in Minkowski space to facilitate future
application to real-time processes.
A Minkowski momentum is denoted $p = (p_0, {\bf p})$,
and the inner product is $p \cdot q = p_0 q_0 - {\bf p} \cdot {\bf q}$.
The vector that specifies the thermal rest frame 
is $n = (1,{\bf 0})$.

\subsection{Gluon self-energy}

The HTL gluon self-energy tensor for a gluon of momentum $p$ is
\bqa
\label{a1}
\Pi^{\mu\nu}(p)=m_D^2\left[
{\cal T}^{\mu\nu}(p,-p)-n^{\mu}n^{\nu}
\right]\;.
\eqa
%
The tensor ${\cal T}^{\mu\nu}(p,q)$, which is defined only for momenta
that satisfy $p+q=0$, is
\bqa
{\cal T}^{\mu\nu}(p,-p)=
\left \langle y^{\mu}y^{\nu}{p\!\cdot\!n\over p\!\cdot\!y}
\right\rangle_{\bf\hat{y}} \;.
\label{T2-def}
\eqa
%
The angular brackets indicate averaging
over the spatial directions of the light-like vector $y=(1,\hat{\bf y})$.
The tensor ${\cal T}^{\mu\nu}$ is symmetric in $\mu$ and $\nu$
and satisfies the ``Ward identity''
\bqa
p_{\mu}{\cal T}^{\mu\nu}(p,-p)=
p\!\cdot\!n\;n^{\nu}\;.
\label{ward-t2}
\eqa
%
The self-energy tensor $\Pi^{\mu\nu}$ is therefore also
symmetric in $\mu$ and $\nu$ and satisfies
\bqa
p_{\mu}\Pi^{\mu\nu}(p)&=&0\;,\\
\label{contr}
g_{\mu\nu}\Pi^{\mu\nu}(p)&=&-m_D^2\;.
\eqa
%

The gluon self-energy tensor can be expressed in terms of two scalar functions,
the transverse and longitudinal self-energies $\Pi_T$ and $\Pi_L$,
defined by
\bqa
\label{pit2}
\Pi_T(p)&=&{1\over d-1}\left(
\delta^{ij}-\hat{p}^i\hat{p}^j
\right)\Pi^{ij}(p)\;, \\
\label{pil2}
\Pi_L(p)&=&-\Pi^{00}(p)\;,
\eqa
%
where ${\bf \hat p}$ is the unit vector
in the direction of ${\bf p}$.
In terms of these functions, the self-energy tensor is
\bqa
\label{pi-def}
\Pi^{\mu\nu}(p) \;=\; - \Pi_T(p) T_p^{\mu\nu}
- {1\over n_p^2} \Pi_L(p) L_p^{\mu\nu}\;,
\eqa
%
where the tensors $T_p$ and $L_p$ are
\bqa
T_p^{\mu\nu}&=&g^{\mu\nu} - {p^{\mu}p^{\nu} \over p^2}
-{n_p^{\mu}n_p^{\nu}\over n_p^2}\;,\\
L_p^{\mu\nu}&=&{n_p^{\mu}n_p^{\nu} \over n_p^2}\;.
\eqa
%
The four-vector $n_p^{\mu}$ is
\bqa
n_p^{\mu} \;=\; n^{\mu} - {n\!\cdot\!p\over p^2} p^{\mu} \, ,
\eqa
%
and satisfies $p\!\cdot\!n_p=0$ and $n^2_p = 1 - (n\!\cdot\!p)^2/p^2$.
The equation~(\ref{contr}) reduces to the identity
\bqa
(d-1)\Pi_T(p)+{1\over n^2_p}\Pi_L(p) \;=\; m_D^2 \;.
\label{PiTL-id}
\eqa
%
We can express both self-energy functions in terms of the function
${\cal T}^{00}$ defined by (\ref{T2-def}):
\bqa
\Pi_T(p)&=& {m_D^2 \over (d-1) n_p^2}
\left[ {\cal T}^{00}(p,-p) - 1 + n_p^2  \right] ,
\label{PiT-T}
\\
\Pi_L(p)&=& m_D^2
\left[ 1- {\cal T}^{00}(p,-p) \right] ,
\label{PiT-L}
\eqa
%

In the tensor ${\cal T}^{\mu \nu}(p,-p)$ defined in~(\ref{T2-def}),
the angular brackets indicate the angular average over
the unit vector $\hat{\bf y}$.
In almost all previous work, the angular average in~(\ref{T2-def}) has been
taken in $d=3$ dimensions. For consistency of higher-order 
corrections, it is essential to take the angular average in $d=3-2\epsilon$
dimensions and analytically continue to $d=3$ only after all poles in
$\epsilon$ have been cancelled.
Expressing the angular average as an integral over the cosine of an angle,
the expression for the $00$ component of the tensor is
\bqa
\!\!\!{\cal T}^{00}(p,-p) \!\! &=& \!\! {w(\epsilon)\over2} \!\!
\int_{-1}^1 \!\! dc\;(1-c^2)^{-\epsilon}{p_0\over p_0-|{\bf p}|c} \, ,
\label{T00-int}
\eqa
%
where the weight function $w(\epsilon)$ is
\bqa
w(\epsilon)={\Gamma(2-2\epsilon)\over\Gamma^2(1-\epsilon)}2^{2\epsilon}
= {\Gamma({3\over2}-\epsilon)
        \over \Gamma({3\over2}) \Gamma(1-\epsilon)} \;.
\label{weight}
\eqa
%
The integral in (\ref{T00-int}) must be defined so that it is analytic
at $p_0=\infty$.
It then has a branch cut running from $p_0=-|{\bf p}|$ to $p_0=+|{\bf p}|$.
If we take the limit $\epsilon\rightarrow 0$, it reduces to
\begin{eqnarray}
{\cal T}^{00}(p,-p) &=&
{p_0 \over 2|{\bf p}|}
                \log {p_0 +|{\bf p}| \over p_0-|{\bf p}|}\;,
\end{eqnarray}
%
which is the expression that
appears in the usual HTL self-energy functions.

\label{app:prop}
\subsection{Gluon propagator}
The Feynman rule for the gluon propagator is
\bqa
i\delta^{ab}\Delta_{\mu\nu}(p) \;,
\eqa
%
where the gluon propagator tensor $\Delta_{\mu\nu}$
depends on the choice of gauge fixing.
We consider two possibilities that introduce an arbitrary
gauge parameter $\xi$:  general covariant gauge and
general Coulomb gauge.
In both cases, the inverse propagator reduces in the
limit $\xi\rightarrow\infty$ to
\bqa
\Delta^{-1}_{\infty}(p)^{\mu\nu}&=&
-p^2 g^{\mu \nu} + p^\mu p^\nu - \Pi^{\mu\nu}(p)\;.
\label{delta-inv:inf0}
\eqa
%
This can also be written
\bqa
\Delta^{-1}_{\infty}(p)^{\mu\nu}&=&
- {1 \over \Delta_T(p)}       T_p^{\mu\nu}
+ {1 \over n_p^2 \Delta_L(p)} L_p^{\mu\nu}\;,
\label{delta-inv:inf}
\eqa
%
where $\Delta_T$ and $\Delta_L$ are the transverse and longitudinal
propagators:
\bqa
\Delta_T(p)&=&{1 \over p^2-\Pi_T(p)}\;,
\label{Delta-T:M}
\\
\Delta_L(p)&=&{1 \over - n_p^2 p^2+\Pi_L(p)}\;.
\label{Delta-L:M}
\eqa
%
The inverse propagator for general $\xi$ is
\bqa
\Delta^{-1}(p)^{\mu\nu}&=&
\Delta^{-1}_{\infty}(p)^{\mu\nu}-{1\over\xi}
p^{\mu}p^{\nu}\hspace{0.2cm}\mbox{covariant}\;,
\label{Delinv:cov}
\\ \nonumber
&=&\Delta^{-1}_{\infty}(p)^{\mu\nu}-{1\over\xi}
\left(p^{\mu}-p\!\cdot\!n\;n^{\mu}\right)
\left(p^{\nu}-p\!\cdot\!n\;n^{\nu}\right)
\\ && 
\hspace{3.7cm}\mbox{Coulomb}\;.
\label{Delinv:C}
\eqa
%
The propagators obtained by inverting the tensors in~(\ref{Delinv:C})
and~(\ref{Delinv:cov}) are
\bqa
\Delta^{\mu\nu}(p)&=&-\Delta_T(p)T_p^{\mu\nu}
+\Delta_L(p)n_p^{\mu}n_p^{\nu}
- \xi {p^{\mu}p^{\nu} \over (p^2)^2}
\nonumber \\
&& \hspace{3.7cm}\mbox{covariant}\;,
\label{D-cov}
\\
&=&-\Delta_T(p)T_p^{\mu\nu}
+\Delta_L(p)n^{\mu}n^{\nu}-\xi{p^{\mu}p^{\nu}\over\left(n_p^2p^2\right)^2}
\nonumber \\
&& \hspace{3.7cm}
\mbox{Coulomb}\;.
\label{D-C}
\eqa
%

It is convenient to define the following combination of propagators:
\bqa
\Delta_X(p) &=& \Delta_L(p)+{1\over n_p^2}\Delta_T(p) \;.
\label{Delta-X}
\eqa
%
Using (\ref{PiTL-id}), (\ref{Delta-T:M}), and (\ref{Delta-L:M}),
it can be expressed in the alternative form
\bqa
\Delta_X(p) &=&
\left[ m_D^2 - d \, \Pi_T(p) \right] \Delta_L(p) \Delta_T(p) \;,
\label{Delta-X:2}
\eqa
%
which shows that it vanishes in the limit $m_D \to 0$.
In the covariant gauge, the propagator tensor can be written
\bqa\nonumber
\Delta^{\mu\nu}(p) &=&
\left[ - \Delta_T(p) g^{\mu \nu} + \Delta_X(p) n^\mu n^\nu \right]
- {n \!\cdot\! p \over p^2} \Delta_X(p)
        \left( p^\mu n^\nu  + n^\mu p^\nu \right)
\\
&&
+ \left[ \Delta_T(p) + {(n \!\cdot\! p)^2 \over p^2} \Delta_X(p)
        - {\xi \over p^2} \right] {p^\mu p^\nu \over p^2} \;. \nonumber \\
\label{gprop-TC}
\eqa
%
This decomposition of the propagator into three terms
has proved to be particularly convenient for explicit calculations.
For example, the first term satisfies the identity
\bqa
\left[- \Delta_T(p) g_{\mu \nu} + \Delta_X(p) n_\mu n_\nu \right]
\Delta^{-1}_{\infty}(p)^{\nu\lambda}  &=&
{g_\mu}^\lambda - {p_\mu p^\lambda \over p^2}
+ {n \!\cdot\! p \over n_p^2 p^2} {\Delta_X(p) \over \Delta_L(p)}
        p_\mu n_p^\lambda \;.
\label{propid:2}
\eqa
%

\subsection{Three-gluon vertex}
\label{app:3gluon}

The three-gluon vertex
for gluons with outgoing momenta $p$, $q$, and $r$,
Lorentz indices $\mu$, $\nu$, and $\lambda$,
and color indices $a$, $b$, and $c$ is
\bqa
i\Gamma_{abc}^{\mu\nu\lambda}(p,q,r)=-gf_{abc}
\Gamma^{\mu\nu\lambda}(p,q,r)\;,
\eqa
%
where $f^{abc}$ are the structure constants and
the three-gluon vertex tensor is
\bqa
\Gamma^{\mu\nu\lambda}(p,q,r)&=&
g^{\mu\nu}(p-q)^{\lambda}+
g^{\nu\lambda}(q-r)^{\mu}
+
g^{\lambda\mu}(r-p)^{\nu}
-m_D^2{\cal T}^{\mu\nu\lambda}(p,q,r)\;.
\label{Gam3}
\eqa
%
The tensor ${\cal T}^{\mu\nu\lambda}$ in the HTL correction term
is defined only for $p+q+r=0$:
\bqa
{\cal T}^{\mu\nu\lambda}(p,q,r) \;=\;
 - \Bigg\langle y^{\mu} y^{\nu} y^{\lambda}
\left( {p\!\cdot\!n\over p\!\cdot\!y\;q\!\cdot\!y}
	- {r\!\cdot\!n\over\!r\cdot\!y\;q\!\cdot\!y} \right)
	\Bigg\rangle\;.
\label{T3-def}
\eqa
%
This tensor is totally symmetric in its three indices and traceless in any
pair of indices: $g_{\mu\nu}{\cal T}^{\mu\nu\lambda}=0$.
It is odd (even) under odd (even) permutations of the momenta $p$, $q$, and
$r$. It satisfies the ``Ward identity''
\bqa
q_{\mu}{\cal T}^{\mu\nu\lambda}(p,q,r) \;=\;
{\cal T}^{\nu\lambda}(p+q,r)-
{\cal T}^{\nu\lambda}(p,r+q)\;.
\label{ward-t3}
\eqa
%
The three-gluon vertex tensor therefore satisfies the Ward identity
\bqa
p_{\mu}\Gamma^{\mu\nu\lambda}(p,q,r) \;=\;
\Delta_{\infty}^{-1}(q)^{\nu\lambda}-\Delta_{\infty}^{-1}(r)^{\nu\lambda}\;.
\label{ward-3}
\eqa
%

\subsection{Four-gluon vertex}
\label{app:4gluon}

The four-gluon vertex
for gluons with outgoing momenta $p$, $q$, $r$, and $s$,
Lorentz indices $\mu$, $\nu$, $\lambda$, and $\sigma$,
and color indices $a$, $b$, $c$, and $d$ is
\bqa
i\Gamma^{\mu\nu\lambda\sigma}_{abcd}(p,q,r,s) &=&
- ig^2\big\{ f_{abx}f_{xcd} \left(g^{\mu\lambda}g^{\nu\sigma}
				-g^{\mu\sigma}g^{\nu\lambda}\right)
\nonumber\\&&\hspace{-2.667cm}
+2m_D^2\mbox{tr}\left[T^a\left(T^bT^cT^d+T^dT^cT^b
\right)\right]{\cal T}^{\mu\nu\lambda\sigma}(p,q,r,s)
\big\}
\nonumber
\\
&&
\hspace{-2.667cm}
+ \; 2 \; \mbox{cyclic permutations}\;,
\eqa
%
where the cyclic permutations are of
$(q,\nu,b)$, $(r,\lambda,c)$, and $(s,\sigma,d)$.
The matrices $T^a$ are the fundamental representation
of the $SU(3)$ algebra with the standard normalization
${\rm tr}(T^a T^b) = {1 \over 2} \delta^{ab}$.
The tensor ${\cal T}^{\mu\nu\lambda\sigma}$
in the HTL correction term is defined only for $p+q+r+s=0$:
\bqa
{\cal T}^{\mu\nu\lambda\sigma}(p,q,r,s) &=&
\Bigg\langle y^{\mu} y^{\nu} y^{\lambda} y^{\sigma}
\left( {p\!\cdot\!n \over p\!\cdot\!y \; q\!\cdot\!y \; (q+r)\!\cdot\!y}
\right.
\nonumber\\&&
\hspace{1cm}
\left.
+{(p+q)\!\cdot\!n\over q\!\cdot\!y\;r\!\cdot\!y\;(r+s)\!\cdot\!y}
+{(p+q+r)\!\cdot\!n\over r\!\cdot\!y\;s\!\cdot\!y\;(s+p)\!\cdot\!y}\right)
\Bigg\rangle\;.
\label{T4-def}
\eqa
%
This tensor is totally symmetric in its four indices and traceless in any
pair of indices: $g_{\mu\nu}{\cal T}^{\mu\nu\lambda\sigma}=0$.
It is even under cyclic or anti-cyclic
permutations of the momenta $p$, $q$, $r$, and $s$.
It satisfies the ``Ward identity''
\bqa\nonumber
q_{\mu}{\cal T}^{\mu\nu\lambda\sigma}(p,q,r,s)&=&
{\cal T}^{\nu\lambda\sigma}(p+q,r,s)
\\ &&
-{\cal T}^{\nu\lambda\sigma}(p,r+q,s)
\label{ward-t4}
\eqa
%
and the ``Bianchi identity''
\bqa\nonumber
{\cal T}^{\mu\nu\lambda\sigma}(p,q,r,s)
+ {\cal T}^{\mu\nu\lambda\sigma}(p,r,s,q)+
{\cal T}^{\mu\nu\lambda\sigma}(p,s,q,r)=0\;.
\\ &&
\label{Bianchi}
\eqa
%

When its color indices are traced in pairs, the four-gluon vertex becomes
particularly simple:
\bqa
\delta^{ab} \delta^{cd} i \Gamma_{abcd}^{\mu\nu\lambda\sigma}(p,q,r,s)
&=& -i g^2 N_c (N_c^2-1) \Gamma^{\mu\nu,\lambda\sigma}(p,q,r,s) \;,
\eqa
%
where the color-traced four-gluon vertex tensor is
\bqa
\Gamma^{\mu\nu,\lambda\sigma}(p,q,r,s)&=&
2g^{\mu\nu}g^{\lambda\sigma}
-g^{\mu\lambda}g^{\nu\sigma}
-g^{\mu\sigma}g^{\nu\lambda}
-m_D^2{\cal T}^{\mu\nu\lambda\sigma}(p,s,q,r)\;.
\label{Gam4}
\eqa
%
Note the ordering of the momenta in the arguments of the tensor
${\cal T}^{\mu\nu\lambda\sigma}$, which comes from the use of the
Bianchi identity (\ref{Bianchi}).
The tensor (\ref{Gam4}) is symmetric
under the interchange of $\mu$ and $\nu$,
under the interchange of $\lambda$ and $\sigma$,
and under the interchange of $(\mu,\nu)$ and $(\lambda,\sigma)$.
It is also symmetric under the interchange of $p$ and $q$,
under the interchange of $r$ and $s$,
and under the interchange of $(p,q)$ and $(r,s)$.
It satisfies the Ward identity
\bqa
p_{\mu}\Gamma^{\mu\nu,\lambda\sigma}(p,q,r,s)
&=&\Gamma^{\nu\lambda\sigma}(q,r+p,s)
-\Gamma^{\nu\lambda\sigma}(q,r,s+p)\;.
\label{ward-4}
\eqa
%

\subsection{Ghost propagator and vertex}
\label{app:ghost}

The ghost propagator and the ghost-gluon vertex depend on the gauge.
The Feynman rule for the ghost propagator is
\bqa
&&{i\over p^2}\delta^{ab}      \hspace{2.5cm}\mbox{covariant}\;,
\\
&&{i\over n_p^2 p^2}\delta^{ab} \hspace{2cm}\mbox{Coulomb}\;.
\eqa
%
The Feynman rule for the vertex in which a gluon with
indices $\mu$ and $a$ interacts with an outgoing ghost
with outgoing momentum $r$ and color index $c$ is
\bqa
&&-gf^{abc}r^{\mu} \hspace{3.3cm}\mbox{covariant}\;,
\\
&&-gf^{abc}\left(r^{\mu}-r\!\cdot\!n\;n^{\mu}\right)
\hspace{1cm}\mbox{Coulomb}\;.
\eqa
%
Every closed ghost loop requires a multiplicative factor of $-1$.

\subsection{HTL counterterm}
\label{app:HTLct}

The Feynman rule for the insertion of an HTL counterterm into a gluon
propagator is
\bqa
-i\delta^{ab}\Pi^{\mu\nu}(p)\;,
\eqa
%
where $\Pi^{\mu\nu}(p)$ is the HTL gluon self-energy tensor given
in~(\ref{pi-def}).

\subsection{Imaginary-time formalism}
\label{app:ITF}

In the imaginary-time formalism,
Minkoswski energies have discrete imaginary values
$p_0 = i (2 \pi n T)$
and integrals over Minkowski space are replaced by sum-integrals over
Euclidean vectors $(2 \pi n T, {\bf p})$.
We will use the notation $P=(P_0,{\bf p})$ for Euclidean momenta.
The magnitude of the spatial momentum will be denoted $p = |{\bf p}|$,
and should not be confused with a Minkowski vector.
The inner product of two Euclidean vectors is
$P \cdot Q = P_0 Q_0 + {\bf p} \cdot {\bf q}$.
The vector that specifies the thermal rest frame
remains $n = (1,{\bf 0})$.

The Feynman rules for Minkowski space given above can be easily
adapted to Euclidean space.  The Euclidean tensor in a given
Feynman rule is obtained from the corresponding Minkowski tensor
with raised indices by replacing each Minkowski energy $p_0$
by $iP_0$, where $P_0$ is the corresponding Euclidean energy,
and multipying by $-i$ for every $0$ index.
This prescription transforms $p=(p_0,{\bf p})$ into $P=(P_0,{\bf p})$,
$g^{\mu \nu}$ into $- \delta^{\mu \nu}$,
and $p\!\cdot\!q$ into $-P\!\cdot\!Q$.
The effect on the HTL tensors defined in (\ref{T2-def}),
(\ref{T3-def}), and (\ref{T4-def}) is equivalent to
substituting $p\!\cdot\!n \to - P\!\cdot\!N$ where $N = (-i,{\bf 0})$,
$p\!\cdot\!y \to -P\!\cdot\!Y$ where $Y = (-i,{\bf \hat y})$,
and $y^\mu \to Y^\mu$.
For example, the Euclidean tensor corresponding to (\ref{T2-def}) is
\bqa
{\cal T}^{\mu\nu}(P,-P)=
\left \langle Y^{\mu}Y^{\nu}{P\!\cdot\!N \over P\!\cdot\!Y}
\right\rangle \;.
\label{T2E-def}
\eqa
%
The average is taken over the directions of the unit vector ${\bf \hat y}$.

Alternatively, one can calculate a diagram
by using the Feynman rules for Minkowski momenta,
reducing the expressions for diagrams to scalars,
and then make the appropriate substitutions,
such as $p^2 \to -P^2$, $p \cdot q \to - P \cdot Q$,
and $n \cdot p \to i n \cdot P$.
For example, the propagator functions (\ref{Delta-T:M})
and (\ref{Delta-L:M}) become
\bqa
\Delta_T(P)&=&{-1 \over P^2 + \Pi_T(P)}\;,
\label{Delta-T}
\\
\Delta_L(P)&=&{1 \over p^2+\Pi_L(P)}\;.
\label{Delta-L}
\eqa
%
The expressions for the HTL self-energy functions $\Pi_T(P)$
and $\Pi_L(P)$ are given by
(\ref{PiT-T}) and (\ref{PiT-L}) with $n_p^2$ replaced by
$n_P^2 = p^2/P^2$ and ${\cal T}^{00}(p,-p)$ replaced by
\bqa
{\cal T}_P &=& {w(\epsilon)\over2}
        \int_{-1}^1dc\;(1-c^2)^{-\epsilon}{iP_0\over iP_0-pc} \;.
\label{TP-def}
\eqa
%
Note that this function differs by a sign from the 00 component
of the Euclidean tensor corresponding
to~(\ref{T2-def}):
\bqa
{\cal T}^{00}(P,-P) = - {\cal T}^{00}(p,-p)\bigg|_{p_0 \to iP_0}
                    = - {\cal T}_P \;.
\eqa
%
A more convenient form for calculating sum-integrals
that involve the function ${\cal T}_P$ is
\bqa
{\cal T}_P &=&
        \left\langle {P_0^2 \over P_0^2 + p^2c^2} \right\ranglec \, ,
\label{TP-int}
\eqa
%
where the angular brackets represent an average over $c$ defined by
\begin{equation}
\left\langle f(c) \right\rangle_{\!c} \equiv w(\epsilon) \int_0^1 dc \,
(1-c^2)^{-\epsilon} f(c) \, ,
\label{c-average}
\end{equation}
%
and $w(\epsilon)$ is given in~(\ref{weight}).

\section{Sum-integrals}
\setcounter{equation}{0}
\label{app:sumint}

In the imaginary-time formalism for thermal field theory, 
the 4-momentum $P=(P_0,{\bf p})$ is Euclidean with $P^2=P_0^2+{\bf p}^2$. 
The Euclidean energy $P_0$ has discrete values:
$P_0=2n\pi T$ for bosons, 
where $n$ is an integer. 
Loop diagrams involve sums over $P_0$ and integrals over ${\bf p}$. 
With dimensional regularization, the integral is generalized
to $d = 3-2 \epsilon$ spatial dimensions.
We define the dimensionally regularized sum-integral by
\bqa
  \hbox{$\sum$}\!\!\!\!\!\!\int_{P} &\equiv& 
  \left(\frac{e^{\gamma_E}\mu^2}{4\pi}\right)^\epsilon\;
  T\sum_{P_0=2n\pi T}\:\int {d^{3-2\epsilon}p \over (2 \pi)^{3-2\epsilon}}\;, 
\label{sumint-def}
\eqa
where $3-2\epsilon$ is the dimension of space and $\mu$ is an arbitrary
momentum scale. 
The factor $(e^{\gamma_E}/4\pi)^\epsilon$
is introduced so that, after minimal subtraction 
of the poles in $\epsilon$
due to ultraviolet divergences, $\mu$ coincides 
with the renormalization
scale of the $\overline{\rm MS}$ renormalization scheme.

Below we list the sum-integrals required to complete the three loop
calculation.  We refer to Ref.~\cite{htlpt2} for details concerning
the sum-integral evaluations.\\

\subsection{One-loop sum-integrals}
The simple one-loop sum-integrals required in our calculations
can be derived from the formulas  
\bqa\nonumber
\sumint_{P}{p^{2m}\over(P^2)^n}
\!\!&=&\!\!
\left({\mu\over4\pi T}\right)^{2\epsilon}
{2\Gamma({3\over2}+m-\epsilon)\Gamma(n-{3\over2}-m+\epsilon)
\over\Gamma(n)\Gamma(2-2\epsilon)}
\\\nonumber
&&\;\times\,
\Gamma(1-\epsilon)\zeta(2n-2m-3+2\epsilon)
e^{\epsilon\gamma_E}
\\ &&
\;\;\;\;\times\,T^{4+2m-2n}(2\pi)^{1+2m-2n}\;.
\eqa

The specific bosonic one-loop sum-integrals needed are

\bqa
\sumint_{P}\log P^2&=&-{\pi^2\over45}T^4\;,\\
\sumint_{P}{1\over P^2}
&=&{T^2\over12}
\left({\mu\over4\pi T}\right)^{2\epsilon}
\Bigg[1+\left(
2+2{\zeta^{\prime}(-1)\over\zeta(-1)}
\right)\epsilon
+{\cal O}(\epsilon^2)
\Bigg]\;, 
\\ 
\sumint_P {1 \over (P^2)^2} &=&
{1 \over (4\pi)^2} \left({\mu\over4\pi T}\right)^{2\epsilon} 
\left[ {1 \over \epsilon} + 2 \gamma_E
+{\cal O}(\epsilon)
\right] \;,
\\ 
\sumint_P {1 \over p^2 P^2} &=&
{1 \over (4\pi)^2} \left({\mu\over4\pi T}\right)^{2\epsilon} 
2 \left[ {1\over\epsilon} + 2 \gamma_E + 2 +{\cal O}(\epsilon)
\right]\;.
 \label{ex1}
\eqa

The number $\gamma_1$ is the first Stieltjes gamma constant
defined by the equation
\begin{equation}
\label{zeta}
\zeta(1+z) = {1 \over z} + \gamma_E - \gamma_1 z + O(z^2)\;.
\end{equation}

We also need some more difficult one-loop sum-integrals 
that involve the HTL function 
defined in~(\ref{TP-def}). 
The specific bosonic sum-integrals needed are
\bqa
\sumint_P {1 \over p^4} {\cal T}_P &=&
{1 \over (4\pi)^2} \left({\mu\over4\pi T}\right)^{2\epsilon}
(-1)\left[ 
{1 \over \epsilon} + 2 \gamma_E 
+ 2\log2
+{\cal O}(\epsilon)
\right]
\;,
\label{exa}
\\ 
\sumint_P {1 \over p^2 P^2} {\cal T}_P &=&
{1 \over (4\pi)^2} \left({\mu\over4\pi T}\right)^{2\epsilon}
\bigg[ 
2 \log2 \left({1 \over \epsilon} + 2 \gamma_E \right)
+ 2 \log^2 2 + {\pi^2 \over 3}  
+{\cal O}(\epsilon)
\bigg] \;, \\ \nonumber
\sumint_P {1 \over p^4} ({\cal T}_P)^2 &=&
{1 \over (4\pi)^2} \left({\mu\over4\pi T}\right)^{2\epsilon}
\left( - {2 \over 3} \right)
\left[ (1+ 2 \log 2) \left( {1 \over \epsilon} + 2 \gamma_E \right)
       - {4 \over 3} + {22 \over 3} \log 2 + 2 \log^2 2 
+{\cal O}(\epsilon)
\right]\;.\\
&&
\label{sumint-T:5}
\eqa
%

\subsection{Two-loop sum-integrals}
The simple two-loop sum-integrals that are needed are

\begin{eqnarray}
\sumint_{PQ} {1\over P^2 Q^2 R^2} &=& {\cal O}(\epsilon) \,,
\\
\sumint_{PQ} {1 \over P^2 Q^2 r^2} &=&
{T^2 \over (4 \pi)^2} \left({\mu\over4\pi T}\right)^{4\epsilon}
{1 \over 12}
\left[ {1 \over \epsilon} + 10 - 12 \log 2
	+ 4 {\zeta'(-1) \over \zeta(-1)} 
+{\cal O}(\epsilon)\right]    \,,
\label{sumint2:2}
\\
\sumint_{PQ} {q^2 \over P^2 Q^2 r^4} &=&
{T^2 \over (4 \pi)^2} \left({\mu\over4\pi T}\right)^{4\epsilon}
{1 \over 6}
\left[ {1 \over \epsilon} + {8 \over 3} + 2 \gamma_E
	+ 2 {\zeta'(-1) \over \zeta(-1)}+{\cal O}(\epsilon) \right]    \,,
\label{sumint2:3}
\\
\sumint_{PQ} {q^2 \over P^2 Q^2 r^2 R^2} &=&
{T^2 \over (4 \pi)^2} \left({\mu\over4\pi T}\right)^{4\epsilon}
{1 \over 9}
\left[ {1 \over \epsilon} + 7.521 +{\cal O}(\epsilon)\right]    \,,
\label{sumint2:4}
\\ \nonumber
\sumint_{PQ} {P\!\cdot\!Q \over P^2 Q^2 r^4} &=&
{T^2 \over (4 \pi)^2} \left({\mu\over4\pi T}\right)^{4\epsilon}
\left( -{1 \over 8} \right)
\left[ {1 \over \epsilon} + {2 \over 9} + 4 \log 2 + {8\over3} \gamma_E
	+ {4\over3} {\zeta'(-1) \over \zeta(-1)} +{\cal O}(\epsilon)\right]     \,,
\\
&&
\label{sumint2:5}
\end{eqnarray}
%
where $R = -(P+Q)$ and $r=|{\bf p} + {\bf q}|$.
%
We also need some more difficult two-loop sum-integrals
that involve the functions ${\cal T}_P$ defined in (\ref{TP-def}).
The specific bosonic sum-integrals needed are
\begin{eqnarray}
\nonumber
\sumint_{PQ} {1 \over P^2 Q^2 r^2} {\cal T}_R &=&
{T^2 \over (4 \pi)^2} \left({\mu\over4\pi T}\right)^{4\epsilon}
\left(- {1\over 48} \right)
\left[ {1 \over \epsilon^2}
\right.
\left.
\;+\; \left( 2 - 12 \log2 + 4 {\zeta'(-1) \over \zeta(-1)} \right)
	{1 \over \epsilon}
\right.\\ &&\left.\hspace{4cm}- 19.83 
+{\cal O}(\epsilon)
\right]    \,, 
\label{sumint2:6}
\\ \nonumber
\sumint_{PQ} {q^2 \over P^2 Q^2 r^4} {\cal T}_R &=&
{T^2 \over (4 \pi)^2} \left({\mu\over4\pi T}\right)^{4\epsilon}
\left(- {1\over 576} \right)
\left[ {1 \over \epsilon^2}
\right.
\left.
\;+\; \left( {26\over3} - {24 \over \pi^2} - 92 \log2
	+ 4 {\zeta'(-1) \over \zeta(-1)} \right) {1 \over \epsilon}
\right.\\ &&\left.
\hspace{4cm}
- 477.7 
+{\cal O}(\epsilon)
\right] \,,  
\label{sumint2:7}
\\ \nonumber
\sumint_{PQ} {P\!\cdot\!Q \over P^2 Q^2 r^4} {\cal T}_R &=&
{T^2 \over (4 \pi)^2} \left({\mu\over4\pi T}\right)^{4\epsilon}
\left(- {1\over 96} \right)
\left[ {1 \over \epsilon^2}
\right.
\left.
\;+\; \left( {8 \over \pi^2} + 4 \log2 + 4 {\zeta'(-1) \over \zeta(-1)} \right)
	{1 \over \epsilon}
\right.\\ &&\left.
\hspace{4cm}+ 59.66 
+{\cal O}(\epsilon)
\right] \,.
\label{sumint2:8}
\end{eqnarray}

\subsection{Three-loop sum-integrals}
The three-loop sum-integrals needed are
\bqa\nonumber
\sumint_{PQR}{1\over P^2Q^2R^2(P+Q+R)^2}&=&
{1\over24(4\pi)^2}T^4
\left({\mu\over4\pi T}\right)^{6\epsilon}
\left[{1\over\epsilon}+{91\over15}
+8{\zeta^{\prime}(-1)\over\zeta(-1)}
\right.\\ &&\left.
\hspace{2cm}
-2{\zeta^{\prime}(-3)\over\zeta(-3)}
+{\cal O}(\epsilon)
\right]\;,\\ \nonumber 
\sumint_{PQR}{(P-Q)^4\over P^2Q^2R^4(Q-R)^2(R-P)^2}&=&
{11\over216(4\pi)^2}T^4
\left({\mu\over4\pi T}\right)^{6\epsilon}
\left[{1\over\epsilon}+{73\over22}
+{12\over11}\gamma_E
+{64\over11}{\zeta^{\prime}(-1)\over\zeta(-1)}
\right.\\&&\left.
\hspace{2cm}
-{10\over11}{\zeta^{\prime}(-3)\over\zeta(-3)}
+{\cal O}(\epsilon)
\right]\;.
\eqa
The three-loop sum-integrals were first calculated by Arnold and Zhai
and calculational details can be found in Ref.~\cite{AZ-95}.

\section{Three-dimensional integrals}
Dimensional regularization can be used to
regularize both the ultraviolet divergences and infrared divergences
in 3-dimensional integrals over momenta.
The spatial dimension is generalized to  $d = 3-2\epsilon$ dimensions.
Integrals are evaluated at a value of $d$ for which they converge and then
analytically continued to $d=3$.
We use the integration measure
\begin{equation}
 \int_p\;\equiv\;
  \left(\frac{e^{\gamma_E}\mu^2}{4\pi}\right)^\epsilon\;
\:\int {d^{3-2\epsilon}p \over (2 \pi)^{3-2\epsilon}}\;.
\label{int-def}
\end{equation}

\subsection{One-loop integrals}
The one-loop integral is given by
\bqa\nonumber
I_n&\equiv&\int_p{1\over(p^2+m^2)^n}\\
&=&{1\over8\pi}(e^{\gamma_E}\mu^2)^{\epsilon}
{\Gamma(n-\mbox{$3\over2$}+\epsilon)
\over\Gamma(\mbox{$1\over2$})
\Gamma(n)}m^{3-2n-2\epsilon}
\;.
\eqa
Specifically, we need
\bqa\nonumber
I_0^{\prime}&\equiv&
\int_p\log(p^2+m^2)\\
&=&
-{m^3\over6\pi}\left({\mu\over2m}\right)^{2\epsilon}
\left[
1
+{8\over3}
\epsilon
+{\cal O}\left(\epsilon^2\right)
\right]\;,\\ 
I_1&=&-{m\over4\pi}\left({\mu\over2m}\right)^{2\epsilon}
\left[
1+2
\epsilon
+{\cal O}\left(\epsilon^2\right)
\right]\;,\\
\label{i2}
I_2&=&{1\over8\pi m}\left({\mu\over2m}\right)^{2\epsilon}
\left[1
+{\cal O}\left(\epsilon\right)
\right]
\;.
\eqa
\subsection{Two-loop integrals}
We also need a few two-loop integrals on the form
\bqa
J_n&=&\int_{pq}{1\over p^2+m^2}{1\over(q^2+m^2)^n}
{1\over({\bf p}+{\bf q})^2} \;, \\
K_n&=&\int_{pq}{1\over p^2+m^2}{1\over(q^2+m^2)}
{1\over[({\bf p}+{\bf q})^2]^n} \;.
\eqa
Specifically, we need $J_1$, $J_2$, and $K_1$
which were calculated in Refs.~\cite{BN-96,KZ-96}:
\bqa
J_1&=&
{1\over4(4\pi)^2}\left({\mu\over2m}\right)^{4\epsilon}
\left[
{1\over\epsilon}+2
+{\cal O}(\epsilon)
\right]\;,\\
J_2&=&
{1\over4(4\pi)^2m^2}\left({\mu\over2m}\right)^{4\epsilon}
\left[1+{\cal O}(\epsilon)
\right]\;,\\
K_2&=&
-{1\over8m^2(4\pi)^2}\left({\mu\over2m}\right)^{4\epsilon}
\left[1+{\cal O}(\epsilon)
\right]\;.
\eqa
\subsection{Three-loop integrals}
We also need a number of three-loop integrals. The specific integrals
we need are listed below and were calculated in
Refs.~\cite{BN-96,KZ-96}. They are special cases of more general
integrals defined in Ref.~\cite{broadhurst}.
\bqa
\nonumber
\int_{pqr}{1\over(p^2+m^2)(q^2+m^2)}{1\over r^2({\bf p}+{\bf q}+{\bf r})^2}
&=&-{m\over2(4\pi)^3}\left({\mu\over2m}\right)^{6\epsilon}
\\ &&
\label{sssfirst}
\hspace{-2cm}
\times\left[
{1\over\epsilon}+8
+{\cal O}(\epsilon)\right]\;,\\ 
\int_{pqr}{(r^2+m^2)\over(p^2+m^2)(q^2+m^2)}
{1\over({\bf p}-{\bf q})^2({\bf q}-{\bf r})^2({\bf r}-{\bf p})^2}
&=&{m\over4(4\pi)^3}\left({\mu\over2m}\right)^{6\epsilon}
\nonumber\\ &&
\hspace{-2cm}
\times
\left[
{1\over\epsilon}+8
+{\cal O}(\epsilon)
\right]\;,\\ 
\int_{pqr}{(r^2+m^2)^2\over(p^2+m^2)(q^2+m^2)}
{1\over({\bf p}-{\bf q})^4({\bf q}-{\bf r})^2({\bf r}-{\bf p})^2}
&=&
-{m\over4(4\pi)^3}\left({\mu\over2m}\right)^{6\epsilon}
\nonumber\\ &&
\hspace{-2cm}
\times
\left[
{1\over\epsilon}+6
+{\cal O}(\epsilon)
\right]\;,
\\
\int_{pqr}{1\over(p^2+m^2)(q^2+m^2)(r^2+m^2)}
{1\over({\bf q}-{\bf r})^2({\bf r}-{\bf p})^2}
&=&
{1\over m(4\pi)^3}\left({\mu\over2m}\right)^{6\epsilon}
\nonumber\\ &&
\hspace{-2cm}
\times
\left[
{\pi^2\over12}
+{\cal O}(\epsilon)
\right]\;,\\
\int_{pqr}{1\over(p^2+m^2)(q^2+m^2)}
{1\over({\bf p}-{\bf q})^2({\bf q}-{\bf r})^2({\bf r}-{\bf p})^2}
&=&
-{1\over8m(4\pi)^3}\left({\mu\over2m}\right)^{6\epsilon}
\nonumber\\ &&
\hspace{-2cm}
\times
\left[
{1\over\epsilon}-2
+{\cal O}(\epsilon)
\right]\;, \\
\int_{pqr}{1\over(p^2+m^2)(q^2+m^2)(r^2+m^2)^2}
{1\over({\bf q}-{\bf r})^2({\bf r}-{\bf p})^2}
&=&\nonumber
-{1\over4m^3(4\pi)^3}\left({\mu\over2m}\right)^{6\epsilon}
\nonumber\\ &&
\hspace{-2cm}
\times
\left[
1-{\pi^2\over6}
+{\cal O}(\epsilon)
\right]\;, \\ \nonumber
\int_{pqr}{1\over(p^2+m^2)(q^2+m^2)[({\bf q}-{\bf r})^2+m^2]
[({\bf r}-{\bf p})^2+m^2]}
&=&-{m\over(4\pi)^3}\left({\mu\over2m}\right)^{6\epsilon}
\\ &&
\hspace{-2cm}
\times
\left[
{1\over\epsilon}+8
-4\log2
+{\cal O}(\epsilon)
\right]\;, \\
\int_{pqr}{1\over(p^2+m^2)(q^2+m^2)[({\bf q}-{\bf r})^2+m^2]
[({\bf r}-{\bf p})^2+m^2]}{({\bf p}-{\bf q})^2\over r^2}
&=&{2m\over(4\pi)^3}\left({\mu\over2m}\right)^{6\epsilon}
\nonumber\\ &&
\hspace{-2cm}
\times
\left[
1-2\log2
+{\cal O}(\epsilon)
\right]\;, \\
\int_{pqr}{1\over(p^2+m^2)(q^2+m^2)[({\bf q}-{\bf r})^2+m^2]
[({\bf r}-{\bf p})^2+m^2]}{({\bf p}-{\bf q})^4\over r^4}
&=&
-{3m\over(4\pi)^3}\left({\mu\over2m}\right)^{6\epsilon}
\nonumber\\ &&
\hspace{-2cm}
\times
\left[
1-{4\over3}\log2
+{\cal O}(\epsilon)
\right]\;, \\
\int_{pqr}{1\over(p^2+m^2)(q^2+m^2)[({\bf q}-{\bf r})^2+m^2]
[({\bf r}-{\bf p})^2+m^2]}{1\over r^2}
&=&
{1\over m(4\pi)^3}\left({\mu\over2m}\right)^{6\epsilon}
\nonumber\\ &&
\hspace{-2cm}
\times
\left[
\log2
+{\cal O}(\epsilon)
\right]\;, \\
\int_{pqr}{1\over(p^2+m^2)(q^2+m^2)[({\bf q}-{\bf r})^2+m^2]
[({\bf r}-{\bf p})^2+m^2]}{({\bf p}-{\bf q})^2\over r^4}
&=&
{1\over3m(4\pi)^3}\left({\mu\over2m}\right)^{6\epsilon}
\nonumber\\ &&
\hspace{-2cm}
\times
\left[
1-\log2+{\cal O}(\epsilon)
\right]\;, \\
\int_{pqr}{1\over(p^2+m^2)(q^2+m^2)[({\bf q}-{\bf r})^2+m^2]
[({\bf r}-{\bf p})^2+m^2]}{1\over r^2({\bf p}-{\bf q})^2}
&=&
{1\over4m^3(4\pi)^3}\left({\mu\over2m}\right)^{6\epsilon}
\nonumber\\ &&
\hspace{-2cm}
\times
\left[
1-\log2
+{\cal O}(\epsilon)
\right]\;, \\ \nonumber
\int_{pqr}{1\over(p^2+m^2)(q^2+m^2)[({\bf q}-{\bf r})^2+m^2]
[({\bf r}-{\bf p})^2+m^2]}{1\over r^4}
&=&
-{1\over24m^3(4\pi)^3}\left({\mu\over2m}\right)^{6\epsilon}
\nonumber\\ &&
\hspace{-2cm}
\times\left[
1+2\log2
+{\cal O}(\epsilon)
\right]\;,
\label{ssslast}
\eqa
Finally, we need the combination 
\bqa\nonumber
\int_{pqr}{1\over(p^2+m^2)(q^2+m^2)(r^2+m^2)}
{({\bf p}-{\bf q})^2\over({\bf q}-{\bf r})^2({\bf r}-{\bf p})^2}
\\
+
\int_{pqr}{(q^2+m^2)\over(p^2+m^2)[({\bf r}-{\bf p})^2+m^2]
[({\bf q}-{\bf r})^2+m^2]}
{1\over r^2({\bf p}-{\bf q})^2} 
&&\nonumber
=
\left({\mu\over2m}\right)^{6\epsilon}
{2m\over(4\pi)^3}\left[1+{\cal O}(\epsilon)\right]\;.
\\ 
\eqa

\bibliographystyle{JHEP}
\providecommand{\href}[2]{#2}\begingroup\raggedright\endgroup

\end{document}